\documentclass[letterpaper]{article}
\usepackage{authblk}
\usepackage[margin=1in]{geometry}

\usepackage{times}  
\usepackage{helvet} 
\usepackage{courier}  
\usepackage[hyphens]{url}  
\usepackage{graphicx} 
\usepackage{amssymb}
\usepackage{stackengine}
\usepackage{xcolor}
\usepackage[round]{natbib}
\usepackage{amsmath}
\usepackage{mathtools}
\usepackage{amsfonts}
\usepackage{bm} 
\usepackage{graphicx}
\usepackage{rotating}
\usepackage{tikz}
\usetikzlibrary{bayesnet}
\usepackage{tabularx,hhline}
\usepackage{array}
\usepackage{tabulary}
\newcolumntype{K}[1]{>{\centering\arraybackslash}p{#1}}
\usepackage{bbm}
\usepackage{array}
\usepackage{multirow}
\usepackage{caption}
\usepackage{subcaption}
\usepackage{mathabx}
\usepackage{tabularx,booktabs} 
\usepackage{standalone}
\usepackage[acronym,smallcaps,nowarn]{glossaries}

\usepackage{times}  
\usepackage{helvet}  
\usepackage{courier}  
\usepackage{url}  
\usepackage{graphicx}  
\makeatletter
\newcommand{\distas}[1]{\mathbin{\overset{#1}{\kern\z@\sim}}}%
\newsavebox{\mybox}\newsavebox{\mysim}
\newcommand{\distras}[1]{%
  \savebox{\mybox}{\hbox{\kern3pt$\scriptstyle#1$\kern3pt}}%
  \savebox{\mysim}{\hbox{$\sim$}}%
  \mathbin{\overset{#1}{\kern\z@\resizebox{\wd\mybox}{\ht\mysim}{$\sim$}}}%
}
\makeatother

\usepackage{hyperref}

\usepackage[
  separate-uncertainty = true,
  multi-part-units = repeat
]{siunitx}
\usepackage{cleveref}
\usepackage[utf8]{inputenc} 
\usepackage[T1]{fontenc}    
\usepackage{hyperref}       
\usepackage{url}            
\usepackage{booktabs}       
\usepackage{amsfonts}       
\usepackage{nicefrac}       
\usepackage{microtype}      

\usepackage{mathabx}
 \usepackage{algorithm,algorithmic}

\usepackage{appendix}

\usepackage{etoolbox}
\makeatletter
\patchcmd{\hyper@makecurrent}{%
    \ifx\Hy@param\Hy@chapterstring
        \let\Hy@param\Hy@chapapp
    \fi
}{%
    \iftoggle{inappendix}{
        \@checkappendixparam{chapter}%
        \@checkappendixparam{section}%
        \@checkappendixparam{subsection}%
        \@checkappendixparam{subsubsection}%
        \@checkappendixparam{paragraph}%
        \@checkappendixparam{subparagraph}%
    }{}%
}{}{\errmessage{failed to patch}}

\newcommand*{\@checkappendixparam}[1]{%
    \def\@checkappendixparamtmp{#1}%
    \ifx\Hy@param\@checkappendixparamtmp
        \let\Hy@param\Hy@appendixstring
    \fi
}
\makeatletter

\newtoggle{inappendix}
\togglefalse{inappendix}

\apptocmd{\appendix}{\toggletrue{inappendix}}{}{\errmessage{failed to patch}}
\apptocmd{\subappendices}{\toggletrue{inappendix}}{}{\errmessage{failed to patch}}
  
\title{The Hawkes Edge Partition Model for Continuous-time Event-based Temporal Networks}

\author[1]{Sikun Yang\thanks{sikunyang@gmail.com}}
\author[2]{Heinz Koeppl\thanks{heinz.koeppl@bcs.tu-darmstadt.de}}
\affil[1]{German Center for Neurodegenerative Disease (DZNE)}
\affil[2]{Department of Electrical Engineering and Information Technology, \protect\\ Technische Universit\"at Darmstadt}
\date{}

\begin{document}

\maketitle

\begin{abstract}
We propose a novel probabilistic framework to model continuous-time interaction events data. Our goal is to infer the \emph{implicit} community structure underlying the temporal interactions among entities, and also to exploit how the community structure influences the interaction dynamics among these nodes. 
To this end, we model the reciprocating interactions between individuals using mutually-exciting Hawkes processes. The base rate of the Hawkes process for each pair of individuals is built upon the latent representations inferred using the hierarchical gamma process edge partition model (HGaP-EPM). 
In particular, our model allows the interaction dynamics between each pair of individuals to be modulated by their respective affiliated communities.
Moreover, our model can flexibly incorporate the auxiliary individuals' attributes, or covariates associated with interaction events. Efficient Gibbs sampling and Expectation-Maximization algorithms are developed to perform inference via P\'olya-Gamma data augmentation strategy. Experimental results on
 real-world datasets demonstrate that our model not only achieves competitive performance for temporal link prediction compared with state-of-the-art methods, but also discovers interpretable latent structure behind the observed temporal interactions.
\end{abstract}
\section{INTRODUCTION}
%
There has been considerable interest in modeling and understanding the information diffusion pathways and interaction dynamics among entities from continuously generated streams of data. These streaming data include the timestamped interaction events among entities (e.g., question-answering threads~\citep{hdhp2017learning}, email communications~\citep{DLS} and interaction events among nations~\citep{BPTD,DPGM,Yang2018,Aaron2019}), and the auxiliary contents created by these interacting entities.  Such temporal interaction data enable us not only to track the \emph{topics} underlying the human-generated contents, but also to understand the network formation and evolving process among these interacting entities.

A fundamental problem in the analysis of continuous-time interaction events is to capture the underlying community structure and reciprocity in these interactions. Reciprocity is a common social norm, in which an individual's actions towards another will increase the likelihood of the same type of action being returned in the near future~\citep{HIRM}. Specifically, Hawkes processes are well-fitted to model such reciprocating behaviors in temporal interactions. To further capture the underlying community structure, some recent works~\citep{HIRM,CTSBM,NetHawkes,DLS,HCCRM} attempt to hybridize statistical models for \emph{static} networks with Hawkes processes to model both \emph{implicit} social structure and \emph{reciprocity} among entities. 
The Hawkes stochastic block models (Hawkes-SBMs)~\citep{HIRM,BHP,CHIP} characterize the interaction dynamics between groups of individuals using mutually-exciting Hawkes processes. 
To further capture the reciprocity between each pair of two 
individuals, 
\citet{HCCRM} proposes to model pair-wise reciprocating dynamics by letting the base intensities depending on the underlying community structure.

%
\begin{figure*}
  \centering
      \includegraphics[width=12.5cm,height=12cm, keepaspectratio]{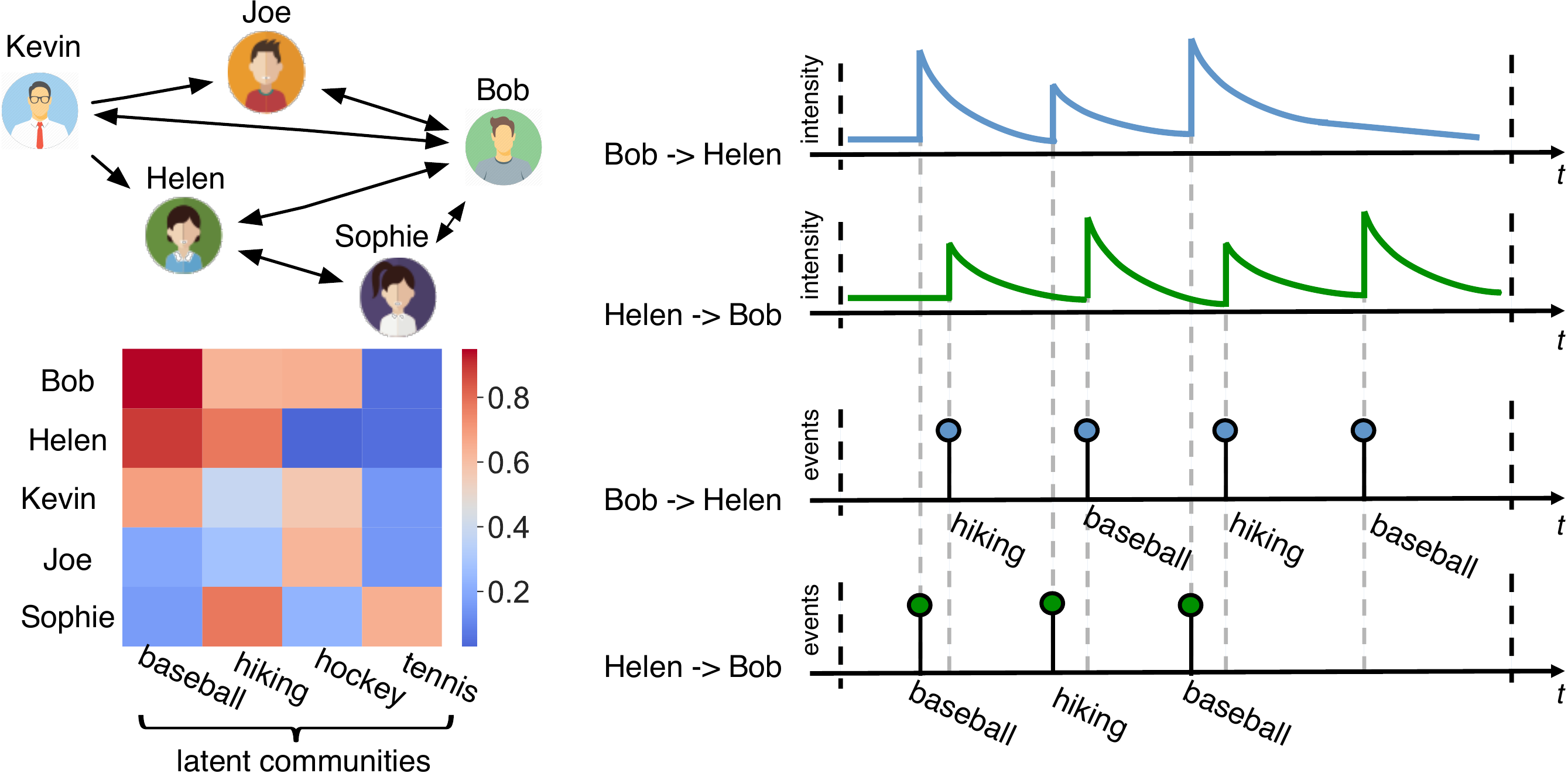}
  \caption{An illustrative example. The top left figure plots the aggregated \emph{directed} networks from the interactions among five nodes. The bottom left graph shows the underlying community structure. We see that both Bob and Helen have interest in ``baseball'' and ``hiking''. The top right graph plots the intensity functions of the interactions from Bob to Helen, and from Helen to Bob, respectively. The bottom right graph plots the interactions from Bob to Helen, and from Helen to Bob, respectively. These interactions may represent the messages between the involved nodes. As in this example, some of their interactions are about ``baseball'', and others relate to ``hiking''. We assume that behind each interaction, the latent patterns of the involved nodes determines the excitation effects of that event on the opposite direction.}
\label{figure_Bob_Helen}
\end{figure*}

Despite having many attractive properties, the Hawkes-CCRM~\citep{HCCRM} is restrictive in that the reciprocity in all the interactions are captured via the same triggering kernel, and thus cannot interpret the differences in interaction dynamics across individuals. 
For example, an employee may reply back to the emails from his/her department more quickly than responding to non-urgent emails from outside. 
A fundamental problem in modeling such temporal dynamics is to infer the \emph{latent struture} behind observed events~\citep{DHP,hdhp2017learning,DMHP,IBPHP}.
To account for heterogeneity both in how two individuals initiate interactions as well as in the dynamics within each specific event, \citet{DLS} proposes to modulate both the \emph{base} and \emph{reciprocate} rate with a dual latent space model, instead of exploiting the latent structure underlying observed events.
 
%

In this paper, we attempt to develop a new framework, the Hawkes edge partition model (Hawkes-EPM) 
, which hybridizes the recently advanced hierarchical gamma process edge partition model (HGaP-EPM)~\citep{EPM} with Hawkes processes. More specifically, the base intensity of the Hawkes process is built upon the latent representations inferred by the HGaP-EPM, which enables us to capture the \emph{overlapping} communities, degree \emph{heterogeneity} and \emph{sparsity} underlying the observed interactions. 
To accurately capture the interaction dynamics between two individuals, our model augments each specific interaction between them with a pair of latent variables, to indicate which of their latent communities (features) leads to the occurring of that interaction. Accordingly, the excitation effect of each interaction on its opposite direction is determined by its latent variables. For instance, as shown in Figure~\ref{figure_Bob_Helen}, Bob and Helen have many common interests (features), and some of their interactions are due to their common interests in playing baseball. 
Moreover, our model can automatically determine the number of the underlying communities via the inherent shrinkage mechanism of the hierarchical gamma process~\citep{Zhou2015}. Furthermore, our model construction can flexibly incorporate the auxiliary individuals' attributes, or covariates associated with interaction events. %
\\
\noindent\textbf{Contributions.} We make the following contributions: (1) We propose a statistical model for continuous-time dynamic networks by capturing the underlying community structure via the base rate of the mutually-exciting Hawkes process, and estimating the number of communities with the hierarchical gamma process. (2) The proposed model accounts for heterogeneity both in \emph{exogenous} and \emph{endogenous} activities. (3) Efficient approximate inference can be performed with closed-form update equations using data augmentation techniques. (4) The developed model is applied for temporal link prediction using real-world data, and shows competitive performance compared with state-of-the-art models.

The paper is organized as follows. Section 2 shortly reviews the necessary background. Section 3 describes the Hawkes-EPM model. Section 4 discusses how the proposed model relates to previous works. Section~5 describes the developed inference procedure. Section~6 presents the experimental results on real-world interaction event datasets.

\section{BACKGROUND}
The proposed Hawkes edge partition model is built upon the hierarchical gamma process edge partition model, which infers the underlying community structure behind the aggregated temporal events, and also relies on Hawkes processes, which capture the reciprocating behaviors between nodes.
Next we shall briefly review the two building components.

\subsection{HIERARCHICAL GAMMA PROCESS EDGE PARTITION MODELS}
The hierarchical gamma process edge partition (HGaP-EPM) model~\citep{EPM} was recently proposed to detect overlapping community structure in static relational data. Formally, let $\mathcal{V}$ denotes a set of nodes, and the (static) relationships among $V\equiv|\mathcal{V}|$ nodes be represented by a binary adjacency matrix $E\in\{0,1\}^{V\times V}$, where $e_{u,v}=1$ if there is an (directed) edge from nodes $u$ to $v$, and $0$ otherwise. We ignore self-edges $\{e_{u,u}\}_{u \in \mathcal{V}}$ as a node never interacts with itself. The (truncated) HGaP-EPM is generated as
\begin{align}
\phi_{u,k} &\sim \mathrm{Gamma}(a_u,1/c_u), \quad a_u \sim \mathrm{Gamma}(e_0,1/f_0), \notag\\
r_k & \sim \mathrm{Gamma}(r_0/K, 1/c_0), \notag\\
\Omega_{k,k'} &\sim \begin{cases} \mathrm{Gamma}(\xi r_{k}, {\chi}), & \text{if}\ k = k' \notag\\ \mathrm{Gamma}(r_{k}r_{k'}, {\chi}), & \text{otherwise}\end{cases},\notag\\\ e_{u,v} &\sim \mathrm{Bernoulli}\Bigg[1-\prod_{k,k'=1}^{K}
\exp(-\phi_{u,k}\Omega_{k,k'}\phi_{v,k'})\Bigg], \notag
\end{align}
where each node $u \in \mathcal{V}$ is chacterized by a \emph{positive} feature vector $[\phi_{u,1},\ldots,\phi_{u,K}]^{\mathrm T}$ with $\phi_{u,k}$ measuring how strongly node $u$ is affiliated to each community $k=1,\ldots,K$. $a_u$ captures the \emph{sociability} of node $u$, and thus node $u$ exhibiting a large number of interactions will be characterized by a large $a_u$. The prevalence of each community $k$ is captured by a positive weight $r_k$. The HGaP-EPM can infer an appropriate number of communities via its inherent shrinkage mechanism: many communities' weights $\{r_k\}$ tend to be small as $K\rightarrow\infty$, and thus most redundant communities will be shrunk effectively. The parameters $\Omega_{k,k}$ and $\Omega_{k,k'}$ capture the intra-community and inter-community interaction weights, respectively. In particular, $\xi$ prevents overly shrinking $\Omega_{k,k}$ for small communities. The probability of there being an edge from node $u$ to node $v$ is parameterized under the Bernoulli-Poisson link (BPL) function $\mathrm{Pr}(y=1 \mid \zeta)=1-e^{-\zeta}$, where $\zeta$ defines the positive rate. 
Following~\citep{EPM}, we impose the $\mathrm{Gamma}(1,1)$ prior over the hyperparameters $c_u, c_0, e_0, f_0, r_0, \xi, \chi$, independently. Interestingly, the probability of an edge $e_{u,v}$ modeled by the BPL can be equivalently generated as
\begin{align}
e_{u,v} & = \mathbf{1}(\tilde{e}_{u,v}\geq1),\notag\\ \tilde{e}_{u,v} & \sim\mathrm{Poisson}\Bigg(\sum_{k=1}^{K}\sum_{k'=1}^{K}\phi_{u,k}\Omega_{k,k'}\phi_{v,k'}\Bigg),\notag
\end{align}
where $\phi_{u,k}\Omega_{k,k'}\phi_{v,k'}$ capture the connecting strength between nodes $u$ and $v$ due to their affiliations to communities $k,k'$, respectively. Note that the HGaP-EPM not only captures the \emph{overlapping} community structure, degree \emph{heterogeneity}, but also characterizes structured sparsity patterns in community-community interactions~\citep{Zhou2018}.
 \subsection{HAWKES PROCESSES}
 Let $N(t)$ be a counting process recording the number of events occurring at times $\{t_i\}$ with $t_i<t$. The probability of an event occurring in a small time interval $[t,t+\mathrm{d}t)$ is given by $\mathrm{Pr}(\mathrm{d}N(t)=1\ |\ \mathcal{H}(t)) = \lambda(t)\mathrm{d}t$, where $\mathcal{H}(t)\equiv\{t_i \mid t_i < t\}$ denotes the history of events up to but not including time $t$, $\mathrm{d}N(t)$ is the increment of the process, and $\lambda(t)$ is the conditional intensity function (intensity, for short) of $N(t)$.
\begin{figure*}[h]
  \centering
      \includegraphics[width=12.5cm,height=10cm, keepaspectratio]{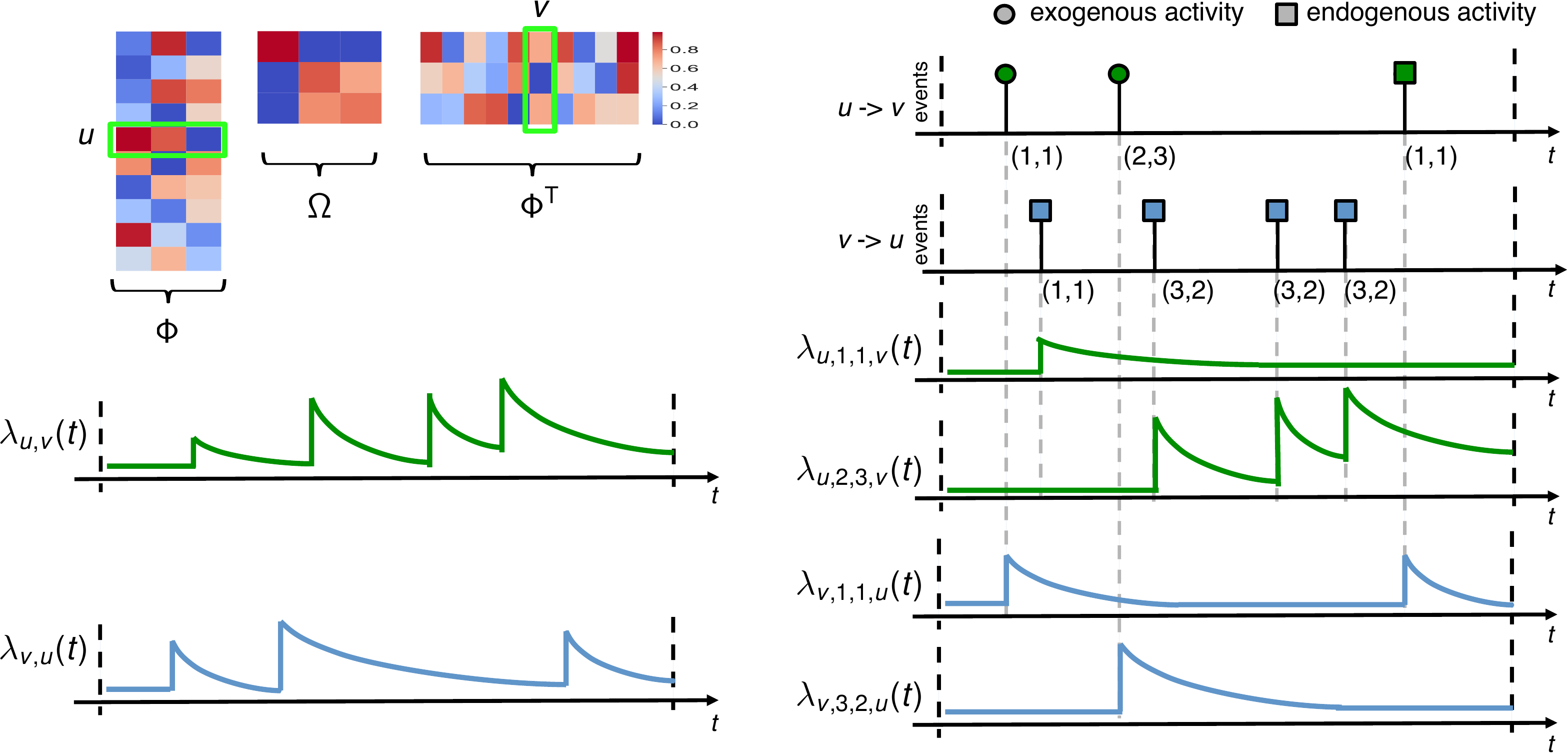}
       \vspace{-0.51em}
  \caption{A simple example for the Hawkes-EPM model. The top left figure shows the inferred matrix of node features $\Phi$, and the community-community interaction strength $\bm{\Omega}$. Here, node $u$ connects to node $v$ through the intra-community interaction $(1,1)$ and inter-community interaction $(2,3)$. The top right figure plots the interaction events between $u$ and $v$. Each event is denoted by a bar, under which we use $(a,b)$ to indicate the latent variables $a, b$ of nodes $u, v$ in that event, e.g., $z_1^u=1, z_1^v=1$ for $1$-st event. The bottom left figure plots the intensities of the interactions from $u$ to $v$, and from $v$ to $u$, respectively. Equivalently, $\lambda_{u,v}(t)$ can be represented by the summation of $\{\lambda_{u,k,k',v}(t)\}_{k,k'}$, where $\lambda_{u,k,k',v}(t)$ denotes the interaction intensity from $u$ to $v$ via the inter-community $(k,k')$.}
\label{HawkesEPM}
\end{figure*}
A Hawkes process is a stochastic point process~\citep{PointProcess} with intensity function defined as
\begin{align}
\lambda(t) = \mu + \int_0^t \gamma(t-s)\mathrm{d}N(s) = \mu + \sum_{j : t_j \in \mathcal{H}(t)} \gamma(t-t_j), \notag
\end{align}
where $\mu\geq0$ is the base rate capturing the \emph{exogenous} activities, and $\gamma(t)$ is the nonnegative triggering kernel modelling the \emph{endogenous} activities. Note that this intensity function characterizes the self-excitation effects that past events have on the current event rate. Here, we consider an exponential kernel $\gamma(t-s)\equiv\alpha \exp[-(t-s)/\delta]$ where $\alpha\geq0$ determines the magnitude of excitations, which exponentially decays with a constant rate $\delta\geq0$.
The stationary condition for Hawkes processes requires $\alpha\delta<1$.
Recent work~\citep{HIRM,DLS,HCCRM,BHP,CHIP} were proposed to capture the reciprocity in communications between a pair of individuals using mutually-exciting Hawkes processes. Formally, for a pair of nodes $u, v \in V$, we have the counting processes $N_{u,v}(t)$, which defines the number of \emph{directed} interactions from node $u$ to node $v$ in the time interval $[0, t)$. Let the history of interactions from nodes $u$ to $v$ be denoted as $\mathcal{H}_{u,v}(t)$. Accordingly, $N_{u,v}(t)$ and $N_{v,u}(t)$ are mutually-exciting Hawkes processes if their intensity functions take the forms
\begin{align}
\lambda_{u,v}(t) & = \mu_{u,v} + \sum_{t_j \in \mathcal{H}_{v,u}(t)} \gamma(t-t_j), \notag\\ 
\lambda_{v,u}(t) & = \mu_{v,u} + \sum_{t_i \in \mathcal{H}_{u,v}(t)} \gamma(t-t_i), \notag
\end{align}
respectively. Note that mutually-exciting Hawkes processes capture the reciprocating interactions from node $u$ to node $v$ at time $t$ as a response to the past interactions from $v$ to $u$.
\section{THE HAWKES EDGE PARTITION MODEL}
Let $\{(t_i, s_i, d_i)\}_{i=1}^{N}$ be a sequence of temporal interaction events, where $(t_i, s_i, d_i)$ is a \emph{directed} interaction from node $s_i$ (sender) to node $d_i$ (receiver) at time $t_i$. 
 To capture reciprocity in interactions, mutually exciting Hawkes processes (MHPs) assume that a specific event $(t_i, s_i, d_i)$ is either an exogenous event triggered by the base rate $\mu_{s_i,d_i}$, or is an endogenous one, responding to a past event. 
 
 To further capture the underlying community structure, we augment each event  $(t_i, s_i, d_i)$ with two auxiliary variables $z_i^s$ and $z_i^d$, which refer to the latent communities affiliated with respectively the sender and receiver. Hence, for the event sequence from node $u$ to node $v$, the first event is driven by one of the sub-rates, $\{\mu_{u,k,k',v}\}_{k,k'}$ where $\mu_{u,k,k',v}$ denotes the sub-rate accounting for the exogenous interactions from $u$ to $v$ due to their respective affiliations to $k, k'$. 
 Accordingly, each subsequent event from $u$ to $v$ is either driven by one of its corresponding sub-rates, or driven by a past event of the opposite direction. Figure~\ref{HawkesEPM} presents a simple illustrative example for the Hawkes Edge Partition Model (Hawkes-EPM).

Formally, for a pair of two nodes $u$ and $v$, the base rate $\mu_{u,v}$ is built upon the latent parameters $\{\phi_{u,k}\}_{u,k}$ and $\{\Omega_{k,k'}\}_{k}$ inferred using the HGaP-EPM.
More specifically, we define the intensity function for nodes $u$ and $v$ as
\begin{equation}\def\useanchorwidth{T}\stackMath
\begin{aligned}
\lambda_{u,v}(t)  &= \sum_{k,k'} \lambda_{u,k,k',v}(t),\label{intensity0}\\
\lambda_{u,k,k',v}(t) &= \mu_{u,k,k',v} + 
  {\stackunder{\sum}{\def\stackalignment{l}%
    {\scriptstyle j : t_j \in \mathcal{H}_{v,k',k,u}(t)}
  }} \gamma_{\scriptstyle k,k'}(t - t_j), \\
\gamma_{\scriptstyle k,k'}(t - s) &= \alpha_{k,k'}\exp[-{(t - s)}/{\delta}], 
\end{aligned}
\end{equation}
where $\lambda_{u,v}(t)$ factorizes into the summation of the sub-intensities $\{\lambda_{u,k,k',v}(t)\}_{k,k'}$. We set the base rate $\mu_{u,v} = \sum_{k,k'}\mu_{u,k,k',v}$, where $\mu_{u,k,k',v}=\phi_{u,k}\Omega_{k,k'}\phi_{v,k'}$. $\phi_{u,k}$ captures the affiliation of node $u$ to community $k$, and $\Omega_{k,k'}$ the inter-community interaction strength between $k$ and $k'$. Hence, the base rate $\mu_{u,v}$ naturally models that two nodes sharing more features are more likely to interact with each other. 

In this work, we assume that if an occurring event is driven by a past event, the latent pattern of the occurring event is also determined by that past event. To this end, the rate $\lambda_{u,k,k',v}(t)$ from $u$ to $v$ under the pattern $(k,k')$, is only allowed to be influenced by the past opposite interactions under the pattern $(k',k)$, $\{(t_j, s_j, d_j) \mid t_j <t, s_j = v, d_j = u, z_j^s = k', z_j^d=k \}$, which we denote by $\{t_j \in\mathcal{H}_{v,k',k,u}(t) \}$. Therefore, in Eq.~(\ref{intensity}), we define a nonnegative kernel function $\gamma_{\scriptstyle k,k'}$, which captures the decaying influence of past events under the pattern $(k',k)$ on the current intensity. More specifically, $\alpha_{k,k'}$ controls the excitatory effect under the pattern $(k', k)$,  and we impose a gamma prior over $\alpha_{k,k'}$, i.e., $\alpha_{k,k'}\sim\mathrm{Gamma}(1,1)$. As reported in related works~\citep{DLS,hdhp2017learning}, we find that inferring time scale $\delta$ suffers from identifiability issue. Instead of modeling temporal dynamics via weighted combinations of basis kernels, we allow $\alpha_{k,k'}$ to be varying between different patterns but fix $\delta$ as a constant. 
Putting all this together, the conditional intensity function of the Hawkes-EPM, for the directed events from $u$ to $v$, is
\begin{equation}\def\useanchorwidth{T}\stackMath
\begin{aligned}
\lambda_{u,v}(t)  &= \mu_{u,v} + 
  {\stackunder{\sum}{\def\stackalignment{l}%
    {\scriptstyle j : t_j \in \mathcal{H}_{v,k',k,u}(t)}
  }} \gamma_{\scriptstyle k,k'}(t - t_j) \label{intensity}\\
& = {\stackunder{\sum}{\def\stackalignment{l}%
                    {\scriptstyle k,k'}
  }} \Bigg\{ \mu_{u,k,k',v} + 
  \mathop{\stackunder{\sum}{\def\stackalignment{l}%
    {\scriptstyle j : t_j \in \mathcal{H}_{v,k',k,u}(t)}
  }} \alpha_{k,k'}\exp[-{(t - t_{j})}/{\delta}] \Bigg\}, 
\end{aligned}
\end{equation}


The latent patterns associated with each interaction is sampled as follows.
If $(t_i, s_i, d_i)$ is an \emph{exogenous} event induced by $\mu_{s_i, d_i}$, the latent patterns $z_i^s, z_i^d$ for $s_i, d_i$ are determined by their affiliated communities via $\bm{\phi}_{s_i}, \bm{\phi}_{d_i}$, respectively. In case that $(t_i, s_i, d_i)$ is an \emph{endogenous} event, $z_i^s, z_i^d$ are determined by the past opposite interactions from $d_i$ to $s_i$. More specifically, the latent patterns associated to $i$-th event can be generated as
\begin{equation}\def\useanchorwidth{T}\stackMath
\begin{aligned}\label{samp_latent_patterns}
&\mathrm{Pr}(z_i^s = k, z_i^d = k' \mid t_i, s_i=u, d_i=v)\\ &= { \Bigg(\mu_{u,k,k',v} +  \mathop{\stackunder{\sum}{\def\stackalignment{l}%
    {\scriptstyle j: t_j \in \mathcal{H}_{v,k',k,u}(t)}
  }} \alpha_{k,k'}\exp[-(t_i - t_{j})/\delta] \Bigg)}/{\lambda_{u,v}(t_i)},\\ &\text{for}\ {\scriptstyle k,\ k'} ={\scriptstyle 1,\ldots,K}. 
\end{aligned}
\end{equation}
In real temporal interactions, some additional information such as auxiliary node attributes, explicitly declared relationships among entities, and communicating contents are also available for accurately modelling temporal interaction dynamics when interaction events are incomplete (say, due to the privacy issues of individuals). Formally, let $\mathbf{x}_{u,v} \equiv [{x}_{u,v}^1, \ldots, {x}_{u,v}^D]^{\mathrm{T}}$ denotes the covariates of $D$-dimension associated with a pair of nodes $u$ and $v$. For example, the covariates $\mathbf{x}_{u,v}$ may represent the common attributes shared by $u$ and $v$, or the word embeddings inferred from the communicating contents between $u$ and $v$. We generalize the 
Hawkes-EPM model by letting 
\begin{align}
\mu_{u,k,k',v} \sim \mathrm{Gamma}(\tilde{\mu}_{u,k,k',v},1/(\exp[- \mathbf{x}_{u,v}^{\mathrm{T}} \bm{\beta}_{k,k'} ])), \label{mu_covdep}
\end{align}
where $\tilde{\mu}_{u,k,k',v} \equiv \phi_{u,k}\Omega_{k,k'}\phi_{v,k'}$, and $\bm{\beta}_{k,k'}\equiv({\beta}_{k,k'}^1, \ldots, {\beta}_{k,k'}^{D})^{\mathrm{T}}$ is the regression coefficient vector of latent pattern $(k,k')$. The base intensity in~(\ref{mu_covdep}) is drawn from a gamma prior where the shape parameter
incorporates the underlying community structure information via $\tilde{\mu}_{u,k,k',v}$, and 
the scale parameter is a function of the input auxiliary covariates.
To our knowledge, the regression component in~(\ref{mu_covdep}) closely relates to~\citep{Piyush15,PBDN}, but firstly applied in this context and the inference derivation is non-trivial.\\
\noindent\textbf{Remarks.} Note that the proposed model allows an unbounded number of latent patterns to be shared across all pairs of interacting nodes via the hierarchical gamma process (HGaP)~\citep{Zhou2015}. As shown in Eq.~(\ref{samp_latent_patterns}), the sub-rate $\mu_{u,k,k',v}$
of the latent pattern $(k, k')$ is non-negligible over the whole time period, and thus our model allows the events widely separated in time but with similar dynamics to be parameterized under the same pattern, to avoid \emph{vanishing} prior issue~\citep{hdhp2017learning,BNHP}.
\section{RELATED WORK}

The proposed model closely relates to the Hawkes process-based interaction models and the Bayesian nonparametric prior-based Hawkes process models.

\noindent\textbf{Hawkes Processes-based Interaction Models.} 
\citet{HIRM} describes the Hawkes stochastic block model (Hawkes-SBM), in which each node is allowed to be affiliated with only one community (non-overlapping), and the interaction dynamics between two nodes are determined by their respective community-specific intensities. The recent extensions~\citep{BHP,CHIP} can be seen as the variants of Hawkes-SBMs. 

\citet{IBPHP} describes an Indian buffet Hawkes process model, which assumes that each event can be simultaneously driven by multiple evolving factors shared among events. In contrast, the Hawkes-EPM relies on a \emph{clustering} structure, where each interaction is categorized as one subtype, while the multiple evolving subtypes are shared among behind the events. 

\citet{HCCRM} describes an unified framework, which captures the overlapping community structure, graph sparsity and degree heterogeneity using compound completely random measure model, and models reciprocity between each pair of nodes via mutually exciting Hawkes processes.
 However, Hawkes-CCRM cannot capture the differences in temporal dynamics of individuals by using the same triggering kernel for all the entities. Our proposed model not only models the interpretable latent structure underlying observed interactions as in~\citep{HCCRM}, but also captures the latent pattern behind each event using community-specific triggering kernels.

\noindent\textbf{Bayesian Nonparametric Hawkes Processes (BNHPs).} 
Recently, Bayesian nonparametric priors (BNPs)~\citep{ferguson1973} are introduced to capture the latent structure underlying the observed event sequence. The Dirichlet-Hawkes process (DHP)~\citep{DHP} models the latent clustering structure underlying the observed events using the Dirichlet process.

The Indian buffet Hawkes process~\citep{IBPHP} and the nested Chinese restaurant process-Hawkes process (NCRP-HP)~\citep{NCRPHP} have been developed to capture the rich factor-structured and hierarchically-structured temporal dynamics, respectively. 

\citet{hdhp2017learning} points out that most previous BNHP models suffer from the vanishing prior problem as the instantiated patterns  in these models are only captured via the endogenous intensity. Hence, an already used pattern will vanish if its intensity tend to be zero. As a consequence, these BNHP methods unavoidably generate many redundant patterns for the events widely separated in time but sharing similar dynamics. 
\citet{hdhp2017learning} resolved this issue using the hierarchical Dirichlet process~\citep{HDP} framework, where the top-layer Dirichlet process defines the distribution over latent patterns, and the bottom-layer Hawkes processes capture the temporal dynamics across multiple event sequences. 
Nevertheless, it is unclear how to generalize the Hierarchical Dirichlet Hawkes Process (HDHP) to model temporal interaction events. Our proposed model infers the appropriate number of communities (patterns) using the hierarchical gamma process prior~\citep{Zhou2015}. 
In the Hawkes-EPM, each latent pattern is modelled by a community-specific intensity function, which is non-negligible over time, and thus effectively prevents from the vanishing prior issue.
\section{INFERENCE}

The proposed model admits efficient approximate inference as the posteriors of all the model parameters are available in closed-form using P\'olya-Gamma data augmentation strategy. Let $\mathcal{D}$ denote the whole events data, $E$ the \emph{binary} adjacency matrix aggregated from $\mathcal{D}$, i.e., ${e_{u,v} = 1}$ for $u,v \in \mathcal{V}$ if there being at least one interaction observed in the time interval $[0,T]$, $\Xi$ the parameters of the HGaP-EPM, and $\Theta$ the parameters of the Hawkes-EPM. The model parameters of the Hawkes-EPM consist of $\{\phi_{u,k},\Omega_{k,k'},\mu_{u,k,k',v},\alpha_{k,k'},\beta_{k,k'},\psi_{k,k'},\omega_{u,k,k',v}\}$. We use the ``$\hat{x}$'' to denote the maximum a posterior (MAP) estimate of $x$. 
A two-step inference procedure is developed to perform maximum-a-posteriori (MAP) estimate: (i) Approximate $\mathrm{Pr}(\Xi \mid \mathcal{D}, E)$ by $\mathrm{Pr}(\Xi \mid E)$, and obtain a MAP estimate $\hat{\Xi}$, and then (ii) Approximate $\mathrm{Pr}(\Theta \mid \Xi, \mathcal{D})$ by $\mathrm{Pr}(\Theta \mid \hat{\Xi}, \mathcal{D})$.
The full posterior is approximated by $\mathrm{Pr}(\Theta, \Xi) = \mathrm{Pr}(\Xi \mid E)\mathrm{Pr}(\Theta \mid \hat{\Xi}, \mathcal{D})$. The posterior inference for $\hat{\Xi}$ is performed using the Gibbs sampling procedure described in~\citep{EPM}. 
Next we shall explain the Expectation-Maximization algorithms to perform MAP estimation following~\citep{EM4HP,ZhouKe13,Xu16}. 
We also use index summation shorthands: $\cdot$ sum out that index, e.g. $x_{\cdot j} = \sum_i x_{ij}$. 
The log-posterior of the observed temporal events $\mathcal{D}\equiv\{(t_i, s_i, d_i)\}_{i=1}^{N}$ is shown in Eq.~(\ref{loglik}).
\begin{figure*}[h]
  \centering
\begin{align} 
\mathcal{L}(\Theta)  = & {\sum}_i \log\Bigg\{ \mu_{s_i,d_i} +  {\sum}_{k,k'}{\sum}_{\scriptstyle j : t_j \in \mathcal{H}_{d_i,k',k, s_i}(t_i) } \alpha_{kk'}\exp\left[-{(t_i - t_{j})}/{\delta}\right] \Bigg\} \notag\\ &- {\sum}_i \left\{\mu_{s_i,d_i}T +  {\sum}_{k,k'}{\sum}_{\scriptstyle j : t_j \in \mathcal{H}_{d_i, k',k,s_i}(t_i)}{\alpha_{kk'}}{\delta}(1 - \exp\left[-{(t_i - t_{j})}/{\delta}\right] ) \right\} + \log \mathrm{Pr}(\Theta). \label{loglik}
\end{align}
\end{figure*}
More specifically, let ${\Theta}^{(l)}$ denote the current model parameters, we construct a tight upper-bound of log-posterior in~(\ref{loglik}) via the Jensen's inequality as
\begin{align} 
& \mathcal{Q}(\Theta \mid \Theta^{(l)}) =\\
& - {\sum}_i \left\{\mu_{s_i,d_i}T +  {\sum}_{k,k'}{\sum}_{\scriptstyle j : t_j \in \mathcal{H}_{d_i,k',k,s_i}(t_i)}\gamma_{kk'}(t_i - t_{j})\right\} \notag\\
& + {\sum}_i {\sum}_{k,k'} \widehat{p}_{i,k,k'} \log\Bigg[ \frac{\mu_{s_i,k,k',d_i}}{\widehat{p}_{i,k,k'}} \Bigg] \notag\\
&+  {\sum}_{k,k'}{\sum}_{\scriptstyle j : t_j \in \mathcal{H}_{d_i,k',k,s_i}(t_i)} \widecheck{p}_{i,k,k'} \Bigg[\frac{\gamma_{kk'}(t_i - t_{j})}{\widecheck{p}_{i,k,k'}} \Bigg] \notag\\&+ \log \mathrm{Pr}(\Theta), \notag
\end{align}
where we define $\widehat{p}_{i,k,k'}$ and $\widecheck{p}_{i,k,k'}$ as
\begin{align}
&\widehat{p}_{i,k,k'} = \frac{\mu^{(l)}_{s_i,k,k',d_i}}{\mu^{(l)}_{s_i,d_i} +  {\sum}_{k,k'}{\sum}_{\scriptstyle j : t_j \in \mathcal{H}_{d_i,k',k,s_i}(t_i)} \gamma^{(l)}_{kk'}(t_i - t_{j})} ,\notag
\end{align}
\begin{align}\label{em_p}
&\widecheck{p}_{i,k,k'} = \frac{{\sum}_{\scriptstyle j : t_j \in \mathcal{H}_{d_i,k',k,s_i}(t_i)} \gamma^{(l)}_{kk'}(t_i - t_{j}) }{\mu^{(l)}_{s_i,d_i} +  {\sum}_{k,k'}{\sum}_{\scriptstyle j : t_j \in \mathcal{H}_{d_i,k',k,s_i}(t_i)} \gamma^{(l)}_{kk'}(t_i - t_{j}) }.
\end{align}
in which $\widehat{p}_{i,k,k'}$ can be interpreted as the probability that $i$-${\mathrm{th}}$ event is drawn from the base rate under the latent pattern $(k, k')$. $\widecheck{p}_{i,k,k'}$ is the probability that $i$-${\mathrm{th}}$ event is triggered by the opposite interaction events under the pattern $(k', k)$.
Accordingly, we update the sufficient statistics as 
\begin{align}
\widehat{m}_{u,k,k',v} &\equiv \sum_{i:s_i=u,d_i=v} \widehat{p}_{i,k,k'},\notag\\
\widecheck{m}_{u,k,k',v} &\equiv \sum_{i:s_i=u,d_i=v} \widecheck{p}_{i,k,k'}.\label{em_m}
\end{align}

Expectations of P\'olya-Gamma random variables are available in closed-form~\citep{EMPG}, and given by
\begin{align}\label{em_omega}
&\mathsf{E}\left[ \omega^{(l+1)}_{u,k,k',v} \right] =\\ &\left(\frac{\tilde{\mu}^{(l)}_{u,k,k',v} + \widehat{m}_{u,k,k',v}}{2\psi^{(l)}_{u,k,k',v}}\right)\tanh\left(\frac{\psi^{(l)}_{u,k,k',v}}{2}\right)\notag
\end{align}
Maximizing $\mathcal{Q}(\Theta)$ with respect to each of the model parameters $\{\mu_{u,k,k',v}\}$, $\{\alpha_{k,k'}\}$, $\{\bm{\beta}_{k,k'}\}$, $\{\bm{\psi}_{k,k'}\}$ fixing the rest, leads to closed-form updates for each of these:
\begin{align}
&\mu_{u,k,k',v}^{(l+1)} = \frac{\tilde{\mu}_{u,k,k',v} + \widehat{m}_{u,k,k',v}}{T + \exp[- \mathbf{x}_{u,v}^{\mathrm{T}} \bm{\beta}^{(l)}_{kk'} ]}. \label{em_mu}
\end{align}

Via the gamma conjugacy, we update $\alpha_{kk'}$ as 
\begin{align}
&\alpha_{k,k'}^{(l+1)}=\notag\\& \frac{e_0 + {\sum}_{u,v}\widecheck{m}_{u,k,k',v}}{f_0 +  {\sum}_i{\sum_{j : t_j \in \mathcal{H}_{d_i,s_i}(t_i)}}{\delta}\left(1 - \exp\left[-\frac{(T - t_{j})}{\delta}\right]\right)}.\label{em_alpha}
\end{align}

Given the expectations of P\'olya-Gamma random variables $\mathsf{E}\left[ \omega_{u,k,k',v} \right]$, we update $\bm{\psi}_{k,k'}$ and $\bm{\beta}_{\scriptscriptstyle k,k'}$ as
\begin{align}
&\bm{\psi}_{k,k'}^{(l+1)} = \left[ \mathrm{diag}( \mathsf{E}\left[ \bm{\omega}^{(l)}_{k,k'} \right]) + \tau\mathbf{I} \right]^{-1} \label{em_psi}\\&\times \left[\frac{\widetilde{m}_{k,k'} - \bm{\mu}^{(l)}_{k,k'}}{2} + \tau(\mathbf{X}^{\mathrm{T}} \bm{\beta}^{(l)}_{k,k'} + \log(T))\right], \notag\\
&\bm{\beta}_{\scriptscriptstyle k,k'}^{(l+1)} = ( \mathbf{X}^{\mathrm{\scriptscriptstyle T}}\mathbf{X} + \tau^{-1}\mathbf{A})^{-1} \mathbf{X}^{\mathrm{T}} \left(\bm{\psi}_{kk'}^{(l)} - \log(T)\right), \label{em_beta}
\end{align}
in which for clarity we define the following notations
\begin{align}
\bm{\omega}^{(l)}_{\scriptscriptstyle k,k'}&\equiv[\omega^{(l)}_{\scriptscriptstyle 1,k,k',1},\ldots,\omega^{(l)}_{\scriptscriptstyle U,k,k',V}], \notag\\
\widetilde{m}_{k,k'}&\equiv [\widehat{m}_{\scriptscriptstyle 1,k,k',1},\ldots,\widehat{m}_{\scriptscriptstyle U,k,k',V}]^{\mathrm{\scriptscriptstyle T}}, \notag\\
\mathbf{A}&\equiv\mathrm{\small diag}[\nu_{\scriptscriptstyle 1}^{\scriptscriptstyle-1},\ldots,\nu_{\scriptscriptstyle D}^{\scriptscriptstyle-1}], \notag\\
\bm{\mu}^{(l)}_{\scriptscriptstyle k,k'}&\equiv[{\mu}^{(l)}_{\scriptscriptstyle 1,k,k',1},\ldots,{\mu}^{(l)}_{\scriptscriptstyle U,k,k',V}]^{\mathrm{\scriptscriptstyle T}}, \notag\\
\mathbf{X}&\equiv[\mathbf{x}_{\scriptscriptstyle 1,1},\ldots,\mathbf{x}_{\scriptscriptstyle U,V}]^{\mathrm{\scriptscriptstyle T}}. \notag
\end{align}

The full procedure of our EM algorithm is summarized in Algorithm~\ref{alg_EM}. We also develop a simple-to-implement Gibbs sampling algorithm, and present its full procedure in the supplement.

\noindent\textbf{Computational Cost.}
For the second inference step, computing the latent variables $\{z_i^s, z_i^d\}$ and updating the intensities for all the given events takes $\mathcal{O}(NK^2)$ time, where $K$ is the estimated number of communities by HGaP-EPM. Estimating $\{ \alpha_{k,k'} \}$ and $\{ \mu_{u,k,k',v} \}$ requires $\mathcal{O}(K^2)$ and $\mathcal{O}(K^2V^2)$ time, respectively. Estimating $\{\beta_{k,k'}\}$ and $\{\psi_{u,k,k',v}\}$ requires solving a linear system, and takes $\mathcal{O}(K^2D^3)$ and $\mathcal{O}(K^2\bar{N})$ time, where $\bar{N}$ denotes the number of node pairs with at least one interaction in $[0, T]$. To sample the P\'olya-Gamma variables $\{\omega_{u,k,k',v}\}$, we employed a fast and accurate approximate sampler of~\citet{SoftplusRegression}, which matches the first two moments of the original distribution. Using the EM algorithm, the P\'olya-Gamma variables are updated in closed-form (as a hyperbolic function)~\citep{EMPG}.

\begin{algorithm}[t]
\caption{Expectation-Maximization algorithm for the Hawkes Edge Partition Model}\label{alg_EM}
\begin{algorithmic}[1]
\REQUIRE events data $\mathcal{D}=\{(t_i,s_i,d_i)\}_{i=1}^N$, $\{\Phi$, $\Omega\}$ inferred by the HGaP-EPM, time scale $\delta$
\ENSURE $\{\mu_{u,k,k',v}\}$, $\{\alpha_{kk'}\}$
\REPEAT 
	\FOR{$n$ = 1:$N$}
	\STATE Update $(\widehat{p}_{i,k,k'},\widecheck{p}_{i,k,k'})$ (Eq.~\ref{em_p})
	\STATE Update the intensity function ${\lambda_{s_i,d_i}(t_i)}$ (Eq.~\ref{intensity})
	\ENDFOR
	\STATE  Update $\widehat{m}_{u,k,k',v}$ and $\widecheck{m}_{u,k,k',v}$ (Eq.~\ref{em_m})
	\STATE  Update the base intensities $\{\mu_{u,k,k',v}\}$ (Eq.~\ref{em_mu})
	\STATE  Update the parameters $\{\bm{\beta}_{kk'}\}$, $\{\omega_{u,k,k',v}\}$, $\{ \psi_{k,k'}\}$ (Eqs.~\ref{em_beta};~\ref{em_omega};~\ref{em_psi})
	\STATE  Update the kernel parameters $\{\alpha_{k,k'}\}$  (Eq.~\ref{em_alpha})
\UNTIL {convergence}
\end{algorithmic}
\end{algorithm}

\section{EXPERIMENTS}
We evaluate the proposed Hawkes-EPM model on three benchmark datasets: 
(1) \textbf{Bosnia.} 
This dataset\footnote{\tiny \url{http://eventdata.parusanalytics.com/data.dir/pevehouse.html}.}
consists of interaction events among 159 nations over 1,819 days (17/01/1991-31/12/1995). There are 1,918 edges, and 34,014 interactions.
(2) \textbf{Gulf.} This dataset\footnote{\tiny \url{http://eventdata.parusanalytics.com/data.dir/gulf.html}.} contains 304,401 interaction events among 202 nations over 7,291 days (15/04/1979-31/03/1999). There are 7,184 edges. 
(3) \textbf{EU-email.} This dataset\footnote{\tiny \url{http://snap.stanford.edu/data/email-EuAll.html}.} consists of 332,334 email communications among 1,005 individuals over 526 days. There are 24,929 edges. We generated the covariate data between each pair of nodes using their common attributes.

We compared our model to two basic models: (1) a Poisson process (PPs) model, which independently models the interaction dynamics between each pair of nodes by a constant event rate, (2) a mutually exciting Hawkes process (MHPs) model, in which we assume the same base rate and kernel parameters for each pair of nodes. Following~\citep{DLS}, we utilized four basis kernels--three exponential kernels with time decaying scale: one hour, one day, one week respectively: $\gamma^{1}(t)\equiv \exp(-24{t}), \gamma^{2}(t)\equiv \exp(-{t}), \gamma^{3}(t)\equiv \exp(-{t}/{7})$, and a periodic kernel $\gamma^{4}(t)\equiv \exp(-{t}/{7})\sin^2({\pi t}/{7})$, and also to three state-of-the-art Hawkes interaction models: (3) the Hawkes Dual Latent Space (DLS) model~\citep{DLS}, which explicitly captures the community structure via the base rate with the Latent space model~\citep{PHoff}, and models the reciprocating dynamics between each pair of nodes via reciprocal latent space model, (4) the Hawkes stochastic block model (Hawkes-SBM), which captures the interaction dynamics using Hawkes process for each community independently, (5) the community Hawkes independent pairs model (CHIP), which models each node pair with a Hawkes process. 
For DLS, we set the latent dimensions $d = 500$ according to the default setting of~\citet{DLS} in our experiments. We demonstrate that the Hawkes-EPM achieves competitive performance but utilizes much fewer latent dimensions ($K_{max} = 100$) compared to DLS. All the baseline models are detailed in the supplementary material. Due to lack of available code, we are not able to compare~\citep{IBPHP,NCRPHP}. 
Given the aggregated graph, we estimated the parameters $\{\bm{\Phi}, \bm{\Omega}\}$ of the HGaP-EPM with the truncation level $K_{\mathrm{max}} = 100$. We ran the Gibbs sampler detailed in~\citep{EPM} for 10,000 MCMC iterations, and used the maximum a posterior estimate $\{\bm{\hat{\Phi}}, \bm{\hat{\Omega}}\}$ in the second step. For the Hawkes-EPM, we choose a kernel decay of $\delta = 1/10$ for the time scale of 10 days. 

\noindent\textbf{Temporal link prediction.} 
To evaluate the predictive performance of all the methods, we sorted the interaction events according to the corresponding timestamps, and made a train-test split so that the training dataset consists of $p$-percent of the whole events with $p$ varying between $50\%$ and $90\%$. 
We trained all the methods using the training datasets. In this task, we let all the models to predict the probability that an edge appears (at least one interaction occurrs) between each pair of nodes in the time interval $[t, t+\hat{\pi})$ with $t$ being the end time of the training events. 
\begin{figure*}
  \centering
   \includegraphics[width=15cm,height=10cm, keepaspectratio]{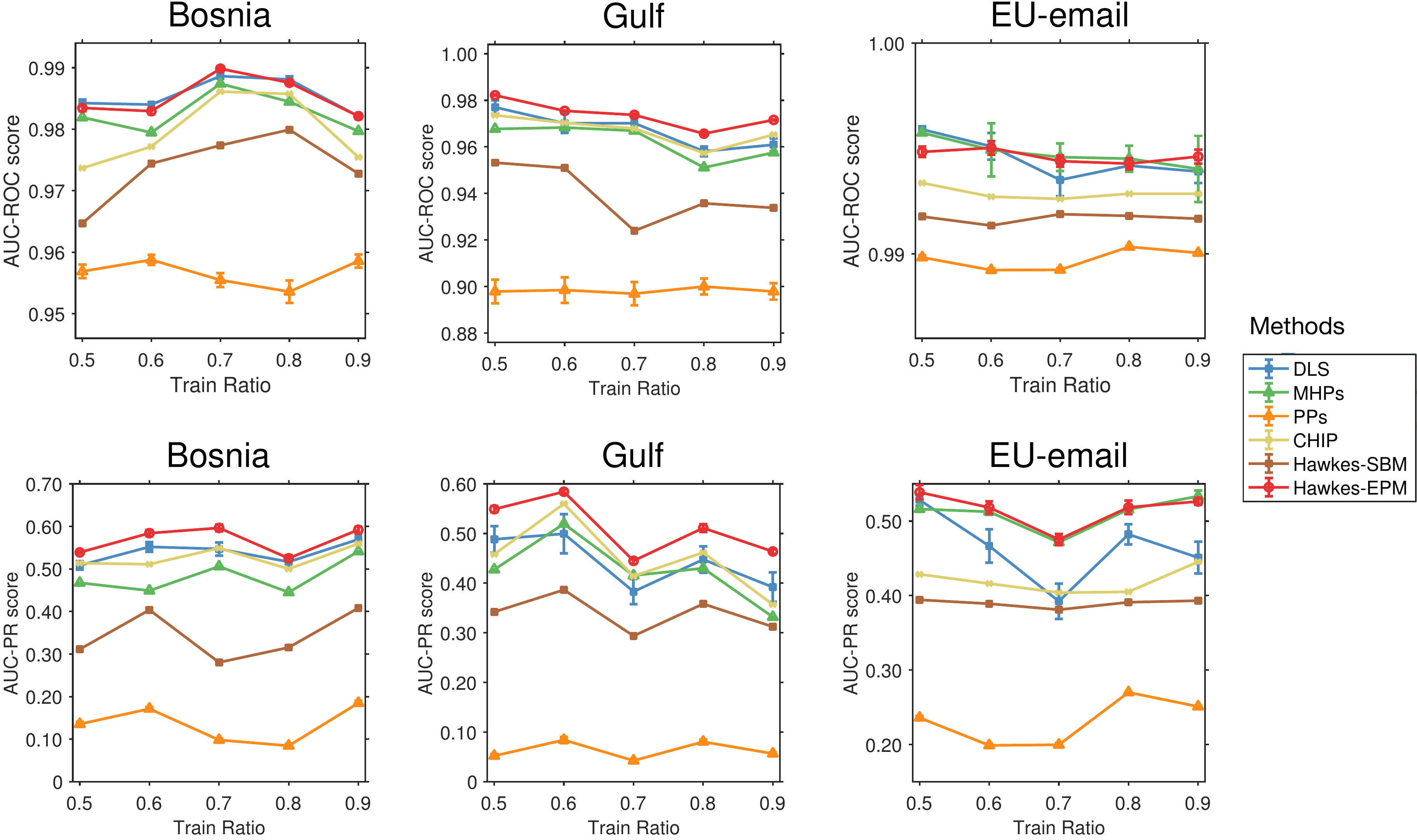}
 \vspace{-0.51em}
  \caption{AUC-ROC and PR scores for the temporal link prediction.}
  \label{fig_roc_pr}
  \end{figure*}
We empirically set $\hat{\pi}$ to be 50 days for all the datasets because it took one or two months for a nation to respond to actions from the other nations on average. 
We calculate the probability of there being at least one interaction in $[t, t+\hat{\pi})$ as $1 - \exp\{-\int_{t}^{t+\hat{\pi}}\lambda_{u,v}(s)\mathrm{d}N_{v,u}(s)\}$. Finally, we compute the average area under the curve (AUC) of both the receiver operating characteristic (ROC) and precision-recall (PR) to evaluate the predictive performance. Although AUC-ROC score is widely used in evaluating link prediction performance, we found that the interactions of the temporal events are quite sparse. Hence, one method can obtain high AUC-ROC score even if it accurately predicts zero-links but shows poor performance in predicting non-zero links. In contrast, AUC-PR score mainly reflects the method's ability to predict non-zero links. 
As shown in Figure~\ref{fig_roc_pr}, the Hawkes process based models (MHPs, DLS, Hawkes-SBM, CHIP, Hawkes-EPM) capture the reciprocating dynamics of the interactions among nodes, and thus significantly outperform the Poisson process model. We noticed that most node pairs exhibit no edges in the time interval $[t,t+\hat{\pi})$ (All the methods accurately predict zero-links and thus achieve high AUC-ROC scores). 
The Hawkes-SBM captures the interaction dynamics of each node pair within the same community only using a single point process, and thus achieves lower predictive scores. It is not surprising that mutually exciting Hawkes processes (MHPs) achieve higher scores as MHPs model each node pair with three periodic kernels, that sufficiently capture the interaction dynamics between each node pairs. 
A closer looking into the AUC-PR scores, shows that the Hawkes-EPM performs better than HPs, CHIP and DLS when the training ratio is low.  This is because the Hawkes-EPM shares the kernel parameters among node pairs, and thus performs well even if most node pairs exhibit few interactions.
Although both the Hawkes-EPM and DLS can capture the heterogeneity in base and reciprocal rate, the Hawkes-EPM effectively exploits the latent  structure behind events and thus consistently outperforms DLS.

\noindent\textbf{Exploratory analysis.}

\begin{figure*}[htb!]
  \centering
      \includegraphics[width=15.5cm,height=5cm, keepaspectratio]{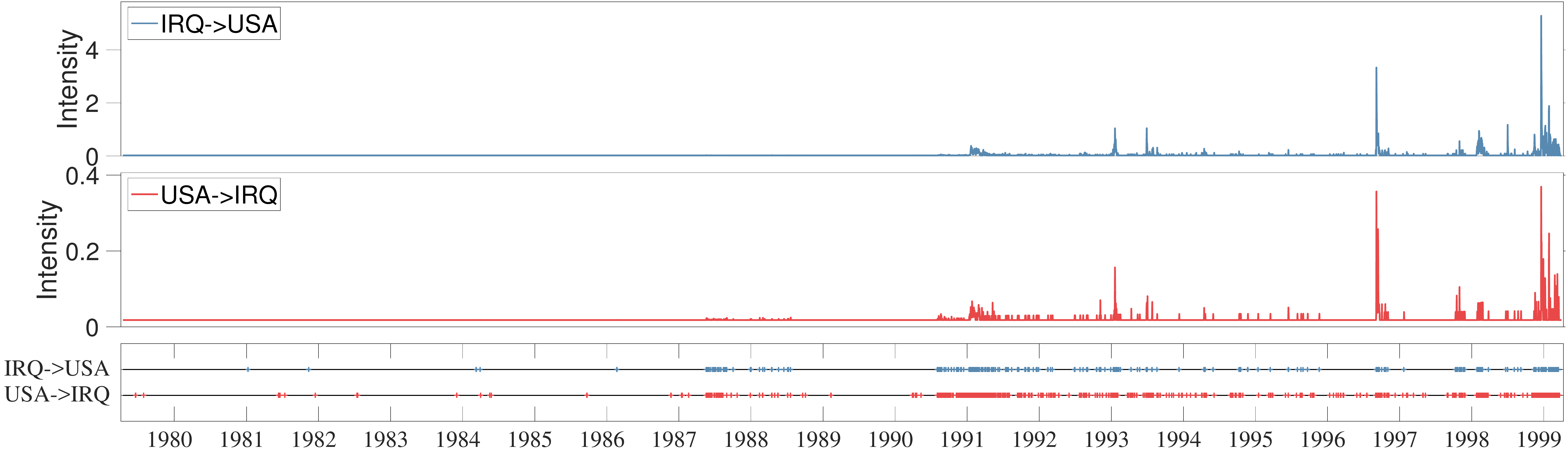}
      \includegraphics[width=15.5cm,height= 5cm, keepaspectratio]{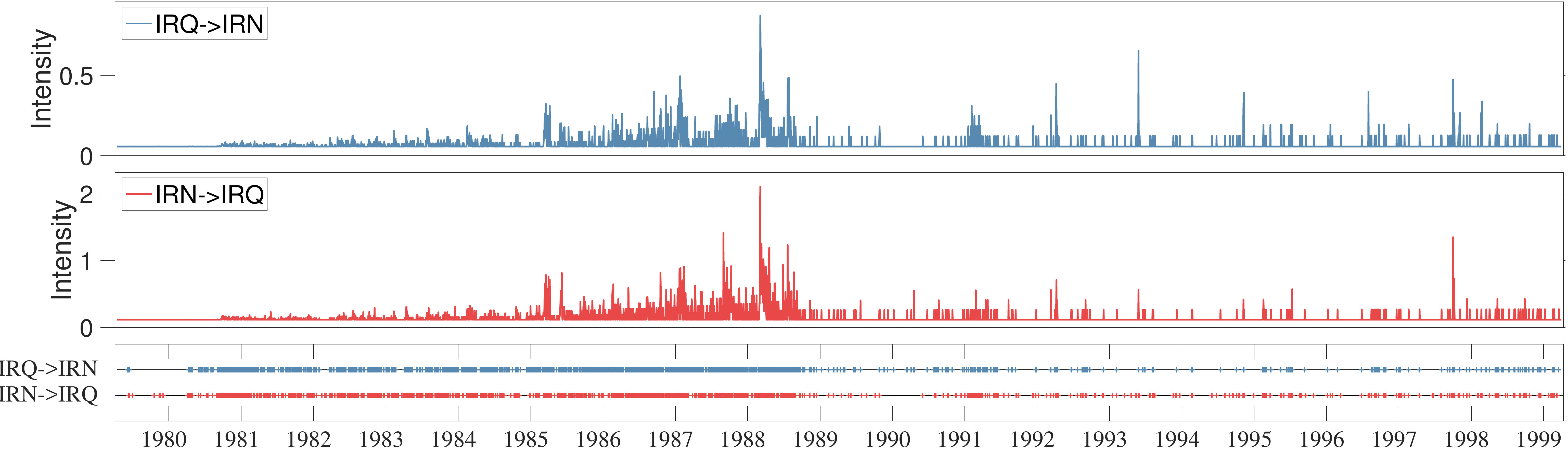}
       \vspace{-0.51em}
  \caption{The top and bottom plot show the intensity of interaction events between USA-Iraq (IRQ), and Iran (IRN)-Iraq (IRQ) inferred by the Hawkes-EPM in the Gulf dataset, respectively.}
  \label{figure_intensity}
\end{figure*}

We also used the Gulf dataset to explore the latent structure estimated by the Hawkes-EPM. We found that $K$ = 12 latent communities, and most of those communities correspond to international military conflicts among nations. Figure~\ref{figure_intensity} shows the inferred intensities of the interaction between USA-Iraq (IRQ), and Iraq (IRQ)-Iran (IRN), respectively. We found that the peaks of the intensities correspond to events surrounding  the Gulf War (1990-1991), the Cruise missile attack on Iraq on 1993 and 1996, the Bombing of Iraq in 1998. In addition, we also plot the intensities of interaction events between Iran (IRN)-Iraq(IRQ). The intensities of the interaction events between these two nodes are gradually increasing from 1980, and reach the peak at 1988. 
To interpret the inferred interaction dynamics between these two nodes, we performed a web search, and found that the Iran-Iraq War started on September, 1980 and ended on August, 1988. Most of the inferred intensities between each pair of nations in the Gulf dataset confirm our knowledge of international affairs. We also provide the additional plots of the intensities between the other nations in the supplementary material.

\section{CONCLUSIONS}

We have presented a probabilistic framework, the Hawkes edge partition model (Hawkes-EPM) for inferring the implicit community structure and reciprocating dynamics among entities from their temporal interactions. The Hawkes-EPM not only models the inherent overlapping communities, sparsity and degree heterogeneity behind interactions, but also captures how the latent communities influence the interaction dynamics among their involved entities.  
Experimental results demonstrate the interpretability and competitive predictive performance of our model in temporal link prediction for several real-world datasets. 
%
Our strategy to cluster events into a set of latent patterns using the gamma process prior~\citep{Zhou2015} combined with Hawkes processes, can be readily generalized to all the closely-related applications, such as continuous-time topic models~\citep{hdhp2017learning} and event-based tensor decomposition~\citep{ShandianZhe2018}.
Another interesting direction is to investigate the privacy-preserving methods for modelling continuously generated events data~\citep{Aaron2018}.
\subsubsection*{Acknowledgements}
We thank the anonymous reviewers for the many useful comments that improved this manuscript.  This research is funded by the European Union's Horizon 2020 research
and innovation programme under grant agreement 668858.
{
\setlength{\bibsep}{0pt plus 0.2ex}
\small
\bibliographystyle{apa}
\bibliography{ref}
}
\clearpage

%
\appendix
\appendixpage
\section{INFERENCE}
Next we shall explain the Gibbs sampling algorithm to infer the parameters of the Hawkes-EPM.
\subsection{GIBBS SAMPLING}
The conditional intensity function of the Hawkes-EPM, for the directed events from $u$ to $v$, is
\begin{equation}\def\useanchorwidth{T}\stackMath
\begin{aligned}
\lambda_{u,v}(t)  &= \mu_{u,v} + 
  {\stackunder{\sum}{\def\stackalignment{l}%
    {\scriptstyle j : t_j \in \mathcal{H}_{v,k',k,u}(t)}
  }} \gamma_{\scriptstyle k,k'}(t - t_j) \label{intensity}\\
& = {\stackunder{\sum}{\def\stackalignment{l}%
                    {\scriptstyle k,k'}
  }} \Bigg\{ \mu_{u,k,k',v} + 
  \mathop{\stackunder{\sum}{\def\stackalignment{l}%
    {\scriptstyle j : t_j \in \mathcal{H}_{v,k',k,u}(t)}
  }} \alpha_{k,k'}\exp[-{(t - t_{j})}/{\delta}] \Bigg\}. 
\end{aligned}
\end{equation}

\noindent\textbf{Sampling latent variables $\{z_i^s, z_i^d\}_{i=1}^N$:} For each event $(t_i,s_i,d_i)$, we utilize an auxiliary binary variable $b_i$ to denote whether $i$-${\mathrm{th}}$ event is triggered by the base rate (exogenous) or by opposite past interactions (endogenous) as 
\begin{align}\label{eq_b}
(b_i \mid -) \sim \mathrm{Bernoulli}(\mu_{s_i,d_i}/\lambda_{s_i,d_i}(t_i)). 
\end{align}
Then, we sample the latent patterns $(z_i^s, z_i^d)$ for each event as
\begin{align}\label{eq_z}
(z_i^s, z_i^d \mid -)&\sim \begin{cases} \mathrm{Cat}\Big(\frac{\{\mu_{s_i,k,k',d_i}\}_{k,k'=1}^{K}}{\lambda_{s_i,d_i}(t_i)}\Big), & \text{if}\ b_i = 1 \\ \mathrm{Cat}\Big(\frac{\{\widecheck{\lambda}_{s_i,k,k',d_i}(t_i)\}_{k,k'=1}^{K}}{\lambda_{s_i,d_i}(t_i)}\Big), & \text{otherwise}\end{cases} 
\end{align}
where $\mathrm{Cat}(\cdot)$ denotes the categorical distribution, and we define 
\begin{equation} \label{update_intensity}
\def\useanchorwidth{T}\stackMath \widecheck{\lambda}_{s_i,k,k',d_i}(t_i) \equiv \mathop{\stackunder{\sum}{\def\stackalignment{l}%
    {\scriptstyle j: t_j \in \mathcal{H}_{d_i,k',k,s_i}(t)}
  }} \alpha_{kk'}\exp[-\delta(t_i - t_{j})]. 
  \end{equation}
  
 Given the sampled latent variables, we update the sufficient statistics as\\ 
 \begin{align}\label{eq_m}
 \widehat{m}_{u,k,k',v} \equiv \sum_{j} \mathbf{1}(b_j=1, s_j = u, d_j = v, z_j^s=k,z_j^d=k'),\\
 \widecheck{m}_{u,k,k',v} \equiv \sum_{j} \mathbf{1}(b_j=0,s_j = u, d_j = v,z_j^s=k,z_j^d=k').\notag
\end{align}
The log-posterior of the observed temporal events $\mathcal{D}\equiv\{(t_i, s_i, d_i)\}_{i=1}^{N}$ is shown in Eq.~\ref{loglik}
\begin{figure*}[htb]
  \centering
\begin{align} \label{loglik}
\mathcal{L}(\Theta)  = & {\sum}_i \log\Bigg\{ \mu_{s_i,d_i} +  {\sum}_{k,k'}{\sum}_{\scriptstyle j : t_j \in \mathcal{H}_{d_i,k',k, s_i}(t_i)} \alpha_{kk'}\exp\left[-{(t_i - t_{j})}/{\delta}\right] \Bigg\} \\ &- {\sum}_i \left\{\mu_{s_i,d_i}T +  {\sum}_{k,k'}{\sum}_{\scriptstyle j : t_j \in \mathcal{H}_{d_i,k',k,s_i}(t_i)}{\alpha_{kk'}}{\delta}(1 - \exp\left[-{(t_i - t_{j})}/{\delta}\right] ) \right\} \notag\\ &+ \log \mathrm{Pr}(\Theta). \notag
\end{align}
\end{figure*}

\noindent\textbf{Sampling the kernel parameters $\{\alpha_{kk'}\}$:} As we place gamma priors over $\alpha_{kk'}$\\ as $\alpha_{kk'}\sim\mathrm{Gamma}(1,1)$, and thus we have
\begin{align}\label{eq_alpha}
&(\alpha_{kk'} \mid -) \sim \mathrm{Gamma}\left(1 + \widecheck{m}_{\cdot k,k'\cdot}, \right. \\ &\left. {1}/{\left[ 1 + {\sum}_i{\sum_{j : t_j \in \mathcal{H}_{d_i,k',k,s_i}(t_i)}}\frac{1}{\delta}\left(1 - \exp\left[-\frac{(T - t_{j})}{\delta}\right]\right) \right]}\right),\notag
\end{align}
where $\widecheck{m}_{\cdot k,k'\cdot} \equiv {\sum}_{i} \widecheck{m}_{s_i,k,k',d_i}$, and $\widecheck{m}_{\cdot k,k'\cdot}$ denotes the total number of endogenous events associated with the latent pattern $(k,k')$.

\noindent\textbf{Sampling the base intensity $\{\mu_{u,k,k',v}\}$:} As we have gamma prior over $\mu_{u,k,k',v}$ \\as $\mu_{u,k,k',v} \sim \mathrm{Gamma}(\tilde{\mu}_{u,k,k',v},1/(\exp[- \mathbf{x}_{u,v}^{\mathrm{T}} \bm{\beta}_{kk'} ]))$, and thus we have

\begin{align}\label{eq_mu}
(\mu_{u,k,k',v} \mid -) \sim &\mathrm{Gamma}\left( \tilde{\mu}_{u,k,k',v} +\widehat{m}_{u,k,k',v}, {1}/{(T+\exp[-\mathbf{x}_{u,v}^{\mathrm{T}} \bm{\beta}_{kk'}])}\right),
\end{align}

Marginalizing out $\mu_{u,k,k',v}$ from the likelihood leads to 
\begin{align}
 \mathrm{Pr}(\mathcal{D} \mid \mathbf{x}_{u,v}, \bm{\beta}_{kk'})& = \int \mathrm{Pr}(\mathcal{D} \mid \mu_{u,k,k',v}) \mathrm{Pr}(\mu_{u,k,k',v} \mid \mathbf{x}_{u,v}, \bm{\beta}_{kk'})  \mathrm{d}\mu_{u,k,k',v} \notag\\
&\ \propto\ \mathrm{NB}(\widehat{m}_{u,k,k',v} ; \tilde{\mu}_{u,k,k',v}, \sigma[\mathbf{x}_{u,v}^{\mathrm{T}} \bm{\beta}_{kk'} + \log(T) ]  ), \notag
\end{align}
where $\sigma(x) = 1/(1 + \exp(-x))$ denotes the logistic function, and $\mathrm{NB}(\cdot)$ denotes the Negative-Binomial distribution. Using the P\'olya-Gamma data augmentation strategy~\citep{Zhou2012,PG}, we first sample 
\begin{align}\label{eq_omega}
(\omega_{u,k,k',v} \mid-) &\sim \mathrm{PG}({\mu}_{u,k,k',v} + \widehat{m}_{u,k,k',v}, \psi_{u,k,k',v}),\notag\\ ({\psi}_{u,k,k',v}\mid-) &\sim \mathcal{N}({\mu}_{\psi}, {\sigma}_{\psi}),
\end{align}
where $\mathrm{PG}$ denotes a P\'olya-Gamma draw, and where
%
\begin{align}
&\psi_{u,k,k',v} \equiv \mathbf{x}_{uv}^{\mathrm{T}} \bm{\beta}_{kk'} + \log(T\pi_{uv}),\notag\\ &\pi_{uv} \sim \log\mathcal{N}(0,\tau^{-1}) \notag\\
&{\sigma}_{\psi} = [\omega_{u,k,k',v} + \tau]^{-1},\notag\\ &\mu_{\psi} = \sigma_{\psi}\left[({\widehat{m}_{u,k,k',v} - {\mu}_{u,k,k',v}})/{2} + \tau(\mathbf{x}_{uv}^{\mathrm{T}} \bm{\beta}_{kk'} + \log(T))\right], \notag
\end{align}
where $\log\mathcal{N}(\cdot)$ denotes the lognormal distribution.

\noindent\textbf{Sampling the regression coefficients $\{\bm{\beta}_{kk'}\}$:}
Given $\{\bm{\psi}_{kk'}\equiv ({\psi}_{1kk'1}, \ldots, {\psi}_{Ukk'V})\}$, we sample $\{\bm{\beta}_{kk'}\}$ as
\begin{align}\label{eq_beta}
(\bm{\beta}_{k,k'}\mid-) &\sim \mathcal{N}(\bm{\mu}_{\beta}, \bm{\Sigma}_{\beta}),
\end{align}
where
$
\bm{\Sigma}_{\beta} = (\tau \mathbf{X}^{\mathrm{T}}\mathbf{X} + \mathbf{A})^{-1}$, $\mathbf{A}\equiv\mathrm{diag}[\nu_1^{-1},\ldots,\nu_{D}^{-1}]$, $\bm{\mu}_{\beta} = \tau\bm{\Sigma}_{\beta} \mathbf{X}^{\mathrm{T}} \left(\bm{\psi}_{kk'} - \log(T)\right)$, and\\ $\mathbf{X} \equiv [\mathbf{x}_{11},\ldots,\mathbf{x}_{UV}]^{\mathrm{T}}. 
$

The full procedure of our Gibbs sampler is summarized in Algorithm~\ref{alg_GS}.
\begin{algorithm}[htb!]
\caption{Gibbs Sampler for the Hawkes Edge Partition Model}\label{alg_GS}
\begin{algorithmic}[1]
\REQUIRE events data $\mathcal{D}=\{(t_i,s_i,d_i)\}_{i=1}^N$, $\{\Phi$, $\Omega\}$ inferred by the HGaP-EPM, maximum iterations $\mathcal{J}$ 
\ENSURE $\{\mu_{u,k,k',v}\}$, $\{\alpha_{kk'}\}$, $\{(z_i^s, z_i^d)\}$
\FOR{$l$ = 1:$\mathcal{J}$}
	\FOR{$n$ = 1:$N$}
	\STATE Sample $b_i$ (Eq.~\ref{eq_b})
	\STATE Sample the latent variables $(z_i^s, z_i^d)$ (Eq.~\ref{eq_z})
	\STATE Update the intensity function ${\lambda_{u,v}(t_i)}$ (Eq.~\ref{intensity})
	\ENDFOR
	\STATE  Update $\widehat{m}_{u,k,k',v}$ and $\widecheck{m}_{u,k,k',v}$ (Eq.~\ref{eq_m})
	\STATE  Sample the base intensities $\{\mu_{u,k,k',v}\}$ (Eq.~\ref{eq_mu})
	\STATE  Sample the parameters $\{\bm{\beta}_{kk'}\}$, $\{\omega_{u,k,k',v}\}$, $\{ \psi_{u,k,k',v}\}$ (Eqs.~\ref{eq_beta};~\ref{eq_omega})
	\STATE  Sample the kernel parameters $\{\alpha_{k,k'}\}$  (Eq.~\ref{eq_alpha})
\ENDFOR
\end{algorithmic}
\end{algorithm}
\section{BASELINE MODELS}
\noindent\textbf{The Hawkes Edge Partition Model (Hawkes-EPM)}
For each pair of nodes $(u, v)$, $u,v \in \mathcal{V}$, and $u \neq v$,
\begin{align}
\mu_{u,k,k',v} &\sim \mathrm{Gamma}(\tilde{\mu}_{u,k,k',v},1/(\exp[- \mathbf{x}_{u,v}^{\mathrm{T}} \bm{\beta}_{kk'} ])), \notag\\
\tilde{\mu}_{u,k,k',v} &\equiv \phi_{u,k}\Omega_{k,k'}\phi_{v,k'}, \notag\\
\bm{\beta}_{k,k'} &\sim \mathcal{N}(\mathbf{0}, \mathbf{A}),\notag\\
\alpha_{kk'} &\sim\mathrm{Gamma}(e_0,1/f_0),\notag \\
\lambda_{u,v}(t) 
& = {\stackunder{\sum}{\def\stackalignment{l}%
                    {\scriptstyle k,k'}
  }} \Bigg\{ \mu_{u,k,k',v} + 
  \mathop{\stackunder{\sum}{\def\stackalignment{l}%
    {\scriptstyle j : t_j \in \mathcal{H}_{v,k',k,u}(t)}
  }} \alpha_{kk'}\exp[-{(t - t_{j})}/{\delta}] \Bigg\}, \notag\\
N_{uv}(t) &\sim \mathrm{Hawkes\ Process}(\lambda_{uv}(t)), \notag
\end{align}
where $\mathbf{A}\equiv\mathrm{diag}[\nu_1^{-1},\ldots,\nu_{D}^{-1}]$.

\noindent\textbf{The Hawkes Dual Latent Space (DLS)~\citep{DLS}}
For each pair of nodes $(u, v)$, $u,v \in \mathcal{V}$, and $u \neq v$,
\begin{align}
\mathbf{z}_v &\sim \mathcal{N}(\mathbf{0}, \sigma^2 \mathrm{I}_{d\times d}), \notag\\
\bm{\mu}_v &\sim \mathcal{N}(\mathbf{0}, \sigma_{\mu}^2 \mathrm{I}_{d\times d}), \notag\\
\bm{\epsilon}_v^{(b)} &\sim \mathcal{N}(\mathbf{0}, \sigma_{\epsilon}^2 \mathrm{I}_{d\times d}), \notag\\
\mathbf{x}_v^{(b)} &\sim \bm{\mu}_v + \epsilon_v^{(b)}, \notag\\
\lambda_{uv}(t) &= \phi\ e^{-\| \mathbf{z}_u - \mathbf{z}_v \|^2_{2}} + \sum_{j:t_j \in \mathcal{H}_{v,u}(t)} \sum_{b=1}^{B}\beta\ e^{-\| \mathbf{x}^{(b)}_u - \mathbf{x}^{(b)}_v \|^2_{2}}\ \gamma_b(t - t_j), \notag\\
N_{uv}(t) &\sim \mathrm{Hawkes\ Process}(\lambda_{uv}(t)). \notag
\end{align}

\noindent\textbf{The Community Hawkes Independent (CHIP) model}
\begin{align}
c_u &\sim \mathrm{Categorical}({\pi_1,\ldots,\pi_k}),\qquad \forall u\in\mathcal{V}\notag\\
\lambda_{uv}(t) &= \phi_{c_u,c_v} + \sum_{j:t_j \in \mathcal{H}_{v,u}(t)} \alpha_{c_u,c_v}\ \exp\{-(t - t_j)/\beta_{c_u,c_v}\}, \notag\\
N_{uv}(t) &\sim \mathrm{Hawkes\ Process}(\lambda_{uv}(t)). \notag
\end{align}

\noindent\textbf{The Hawkes Stochastic Block (Hawkes-SBM) model}
\begin{align}
c_u &\sim \mathrm{Categorical}({\pi_1,\ldots,\pi_k}),\qquad \forall u\in\mathcal{V}\notag\\
\lambda_{k,k'}(t) &= \phi_{k,k'} + \sum_{j:t_j \in \mathcal{H}_{k',k}(t)} \alpha_{k,k'}\ \exp\{-(t - t_j)/\beta_{k,k'}\}, \notag\\
N_{k,k'}(t) &\sim \mathrm{Hawkes\ Process}(\lambda_{k,k'}(t)). \notag
\end{align}

\noindent\textbf{The Mutually Exciting Hawkes processes (MHPs) model}
\begin{align}
\lambda_{uv}(t) &= \phi + \sum_{j:t_j \in \mathcal{H}_{v,u}(t)} \sum_{b=1}^{B} \beta_b\ \gamma_b(t - t_j), \notag\\
N_{uv}(t) &\sim \mathrm{Hawkes\ Process}(\lambda_{uv}(t)). \notag
\end{align}

\noindent\textbf{Poisson process (PPs) model}
\begin{align}
\lambda_{uv}(t) &= \phi_{uv}, \notag\\
N_{uv}(t) &\sim \mathrm{Poisson\ Process}(\lambda_{uv}(t)). \notag
\end{align}
\clearpage
\section{NUMERICAL SIMULATIONS}
\begin{figure*}[ht]
  \centering
      \includegraphics[ width=10cm,height=15cm,keepaspectratio]{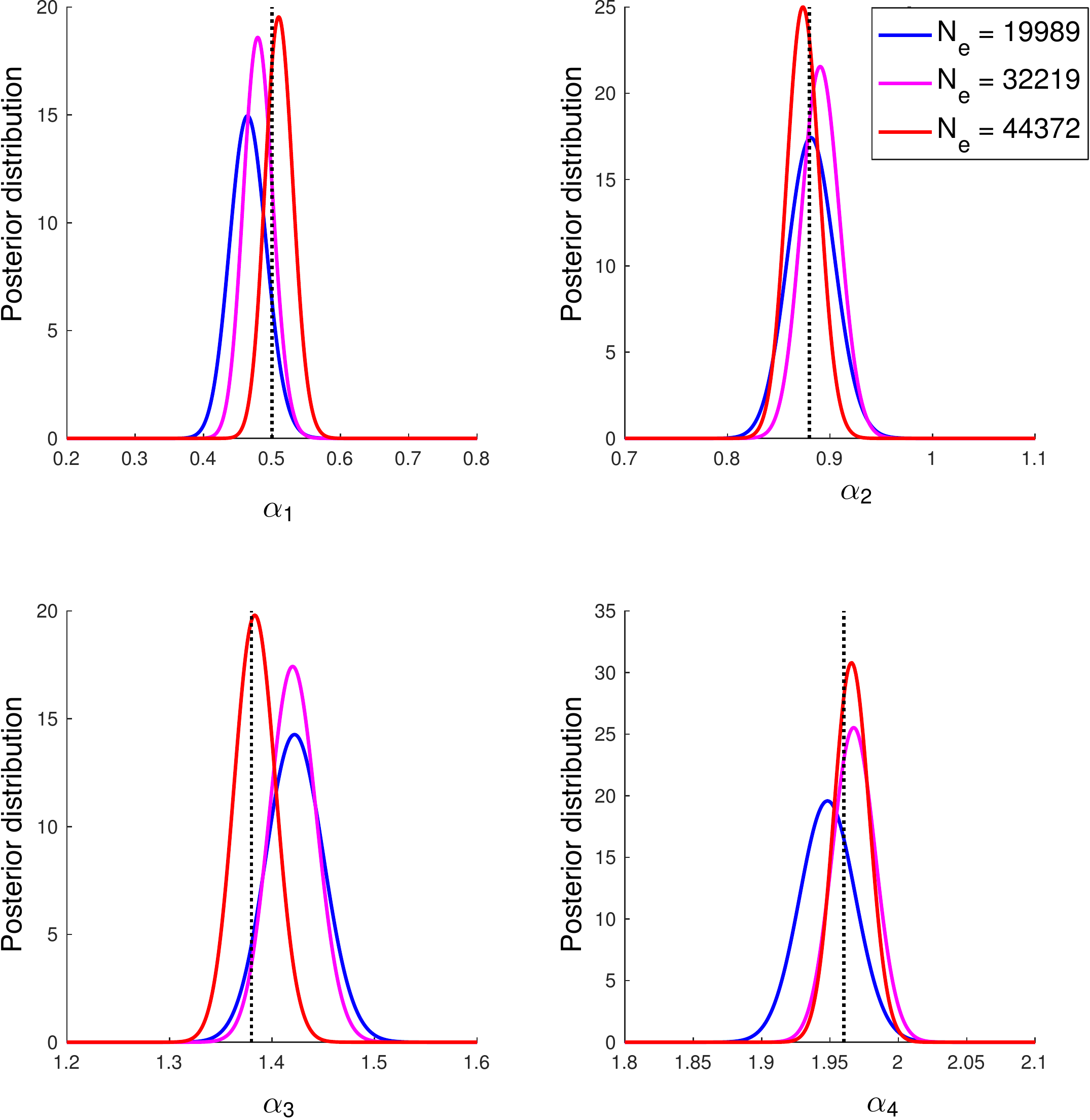}%
  \caption{The posterior distribution of the estimated parameters $\{\alpha_{k}\}$ for the four simulations with the number of events $N_e$. The dashed line indicates the true values of $\{\alpha_{k}\}$.}
  \label{examples1}
\end{figure*}

In this experiment we use synthetic data to evaluate the performance of the Hawkes-EPM in estimating the kernel parameters. 
We consider a collection of nodes $|\mathcal{V}| = 100$, and $K= 4$ latent communities. We generated the base rate $\mu_k\sim \mathrm{Uniform}[0,1]$, and set the kernel parameters $[\alpha_1,\alpha_2,\alpha_3,\alpha_4] = [0.5, 0.88, 1.38, 1.96]$, and $\delta = 0.45$. 
Via the derived Gibbs sampler, the Hawkes-EPM infers the number of latent communities. 
As shown in Figure~(\ref{examples1}), the posterior distributions of the estimated $\{\alpha_k\}$ concentrate toward the true values as the number of observed events is increasing. 
\clearpage
\section{ADDITIONAL RESULTS}
\begin{figure}[htb]
  \centering
      \includegraphics[width=14cm,height=15cm, keepaspectratio]{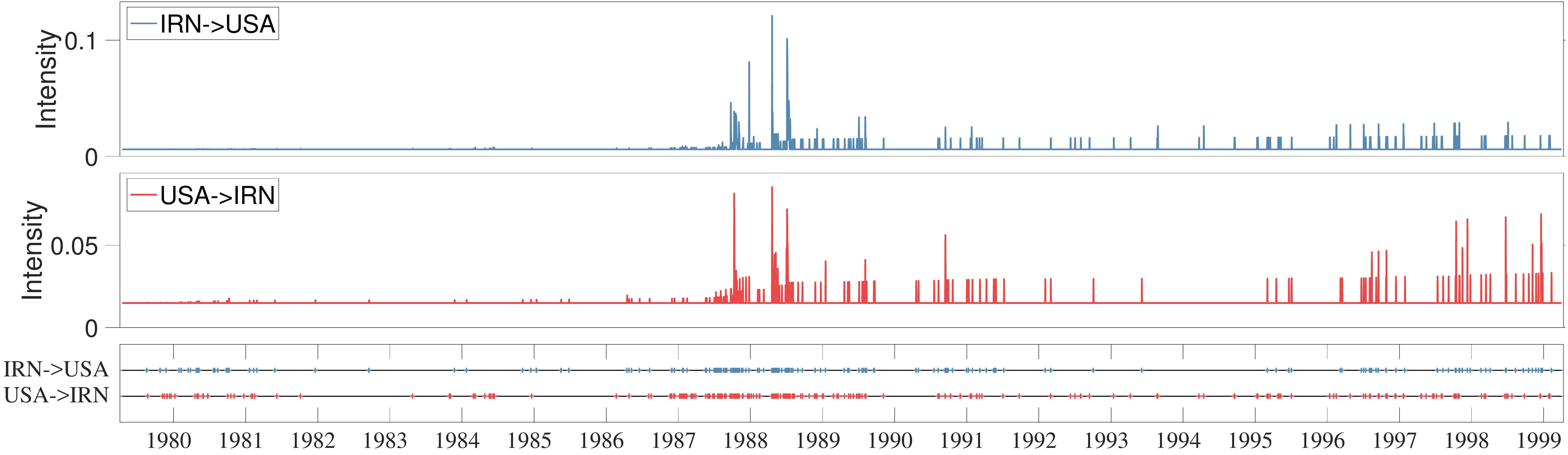}
       \includegraphics[width=14cm,height=15cm, keepaspectratio]{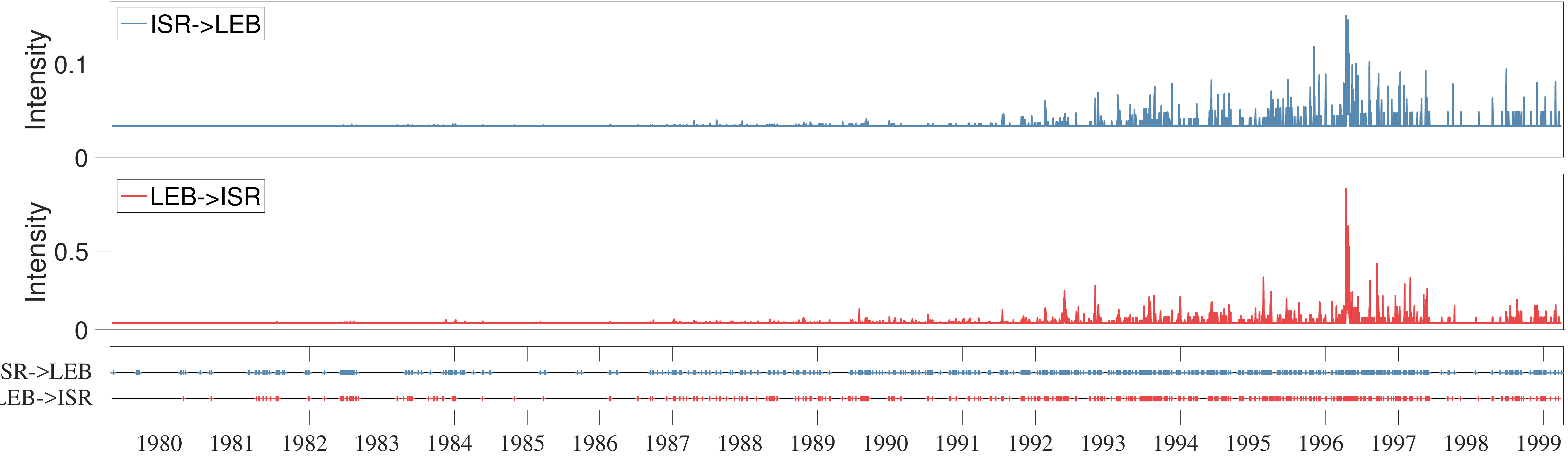}
       \includegraphics[width=14cm,height=15cm, keepaspectratio]{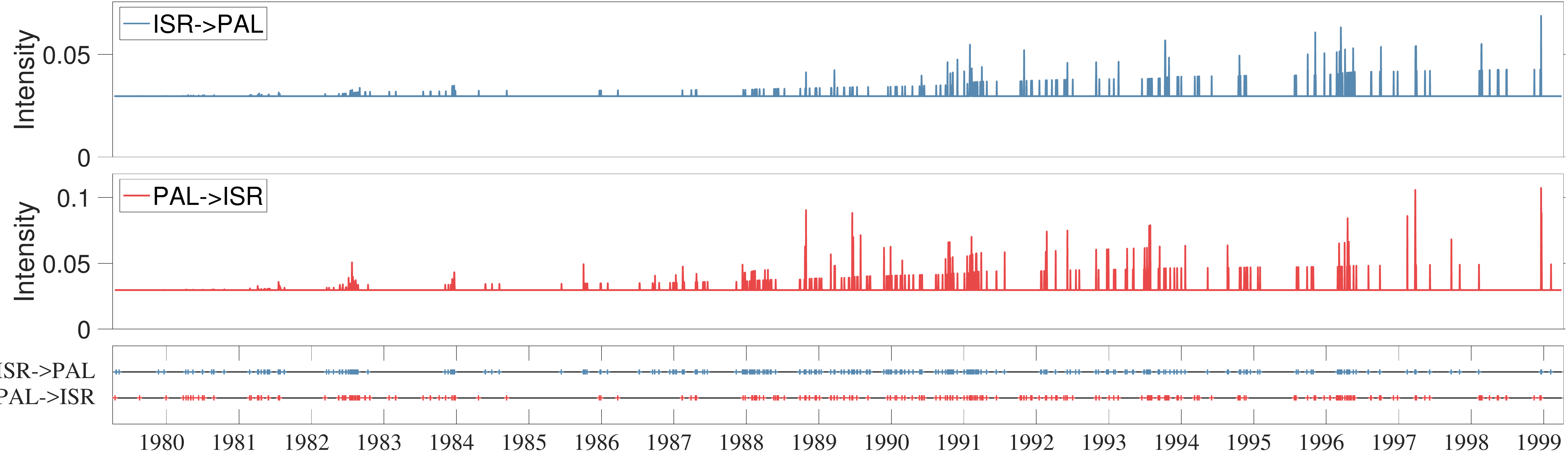}
      \includegraphics[width=14cm,height=15cm, keepaspectratio]{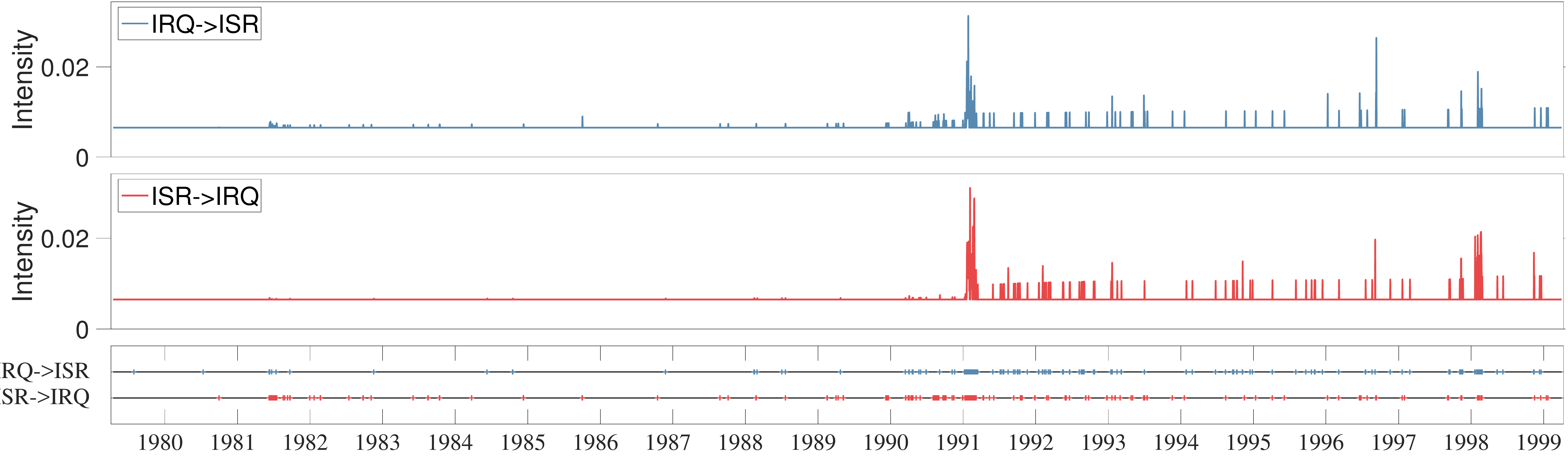}
       \vspace{-0.51em}
  \caption{The plots show the intensity of interaction events among nations inferred by the Hawkes-EPM in the Gulf dataset.}
  \label{figure_intensity21}
\end{figure}

\begin{figure}
  \centering
      \includegraphics[width=14cm,height=15cm, keepaspectratio]{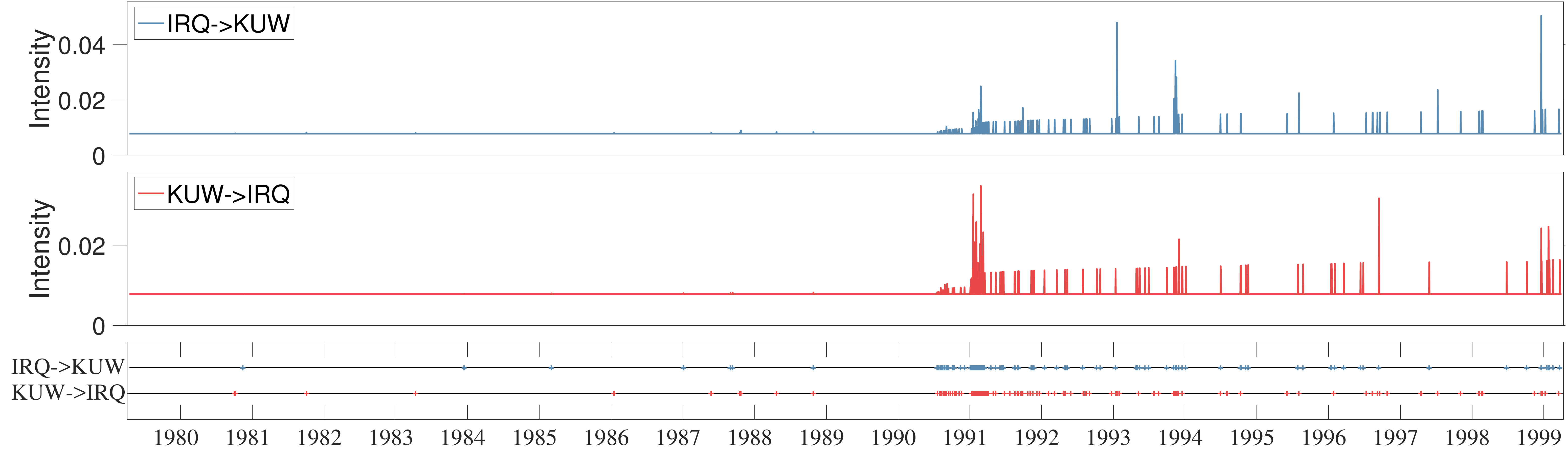}
      \includegraphics[width=14cm,height=15cm, keepaspectratio]{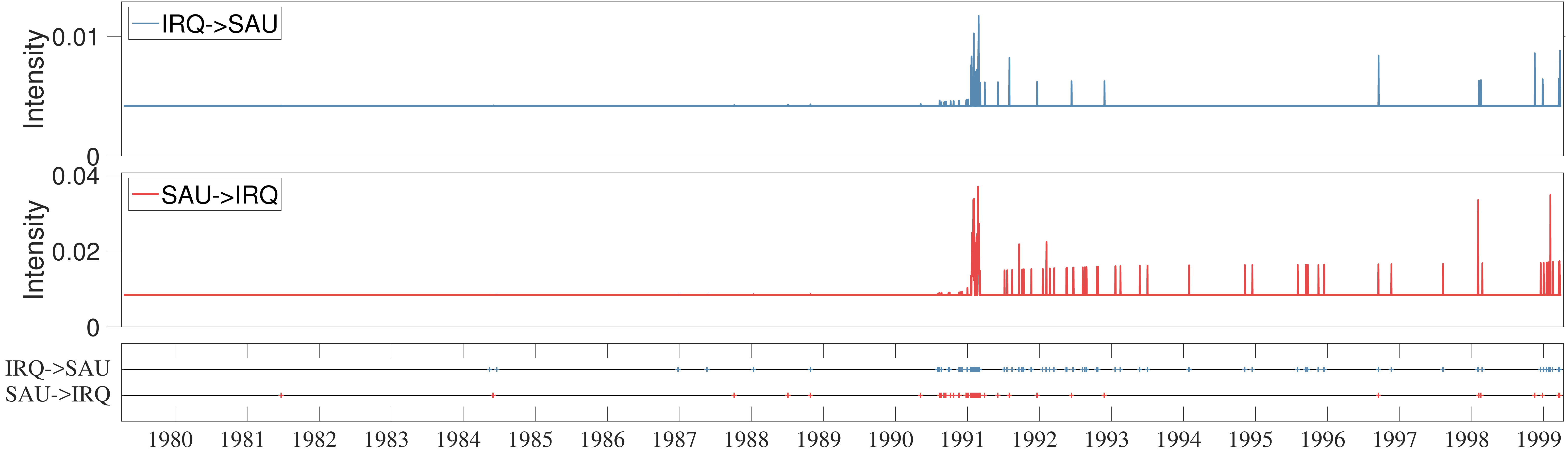}
      \includegraphics[width=14cm,height=15cm, keepaspectratio]{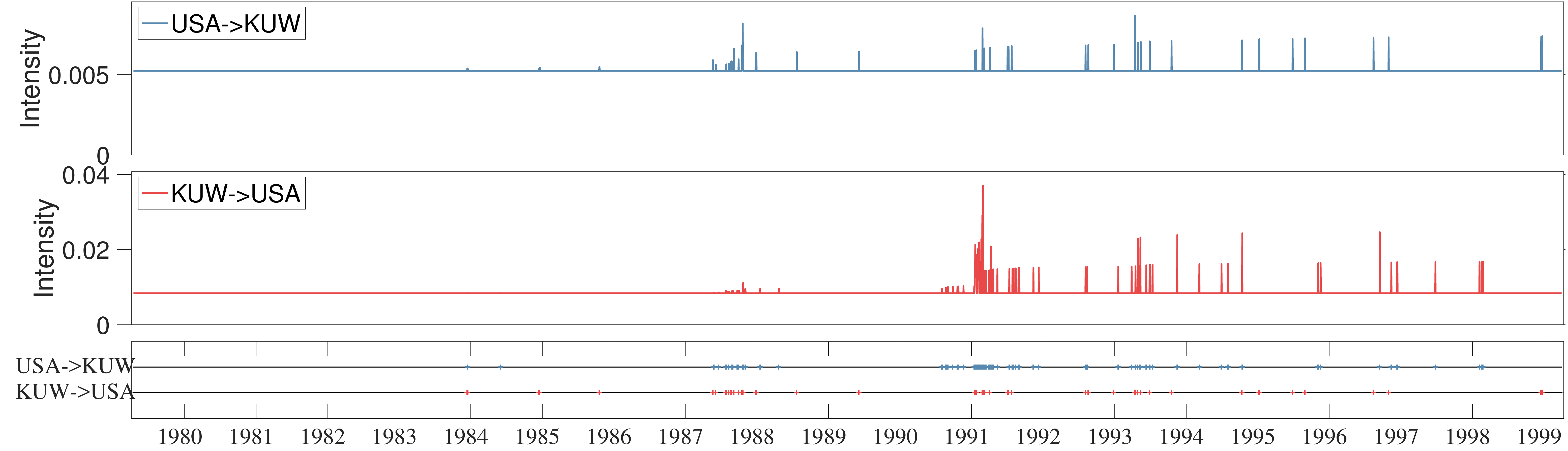}
      \includegraphics[width=14cm,height=15cm, keepaspectratio]{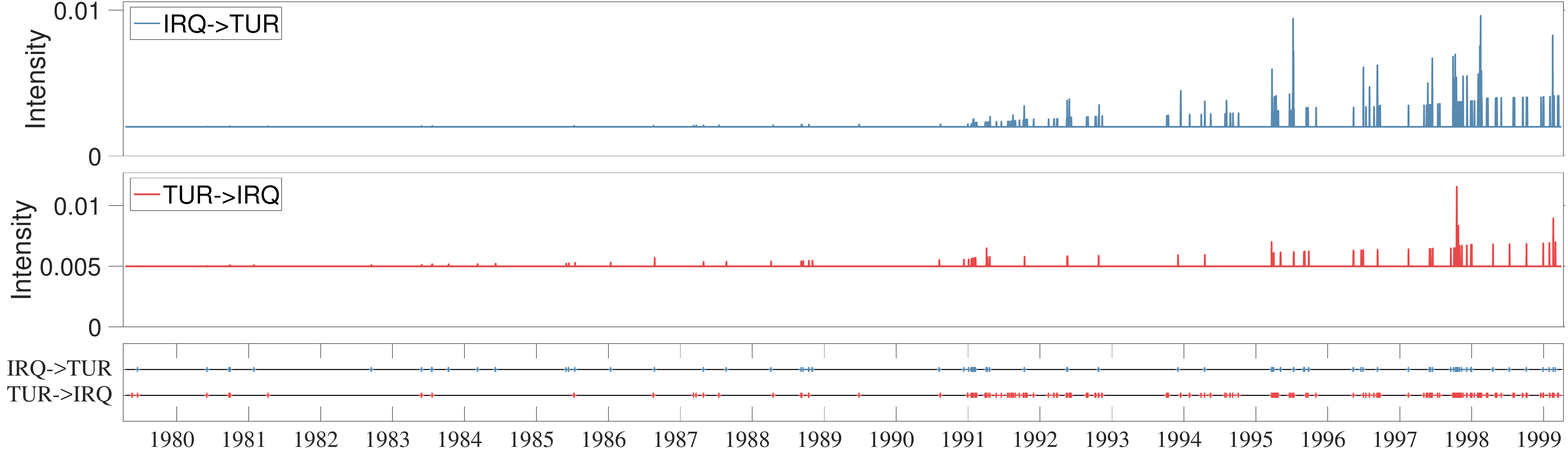}
       \vspace{-0.51em}
  \caption{The plots show the intensity of interaction events among nations inferred by the Hawkes-EPM in the Gulf dataset.}
  \label{figure_intensity22}
\end{figure}

\begin{figure}
  \centering
      \includegraphics[width=14cm,height=15cm, keepaspectratio]{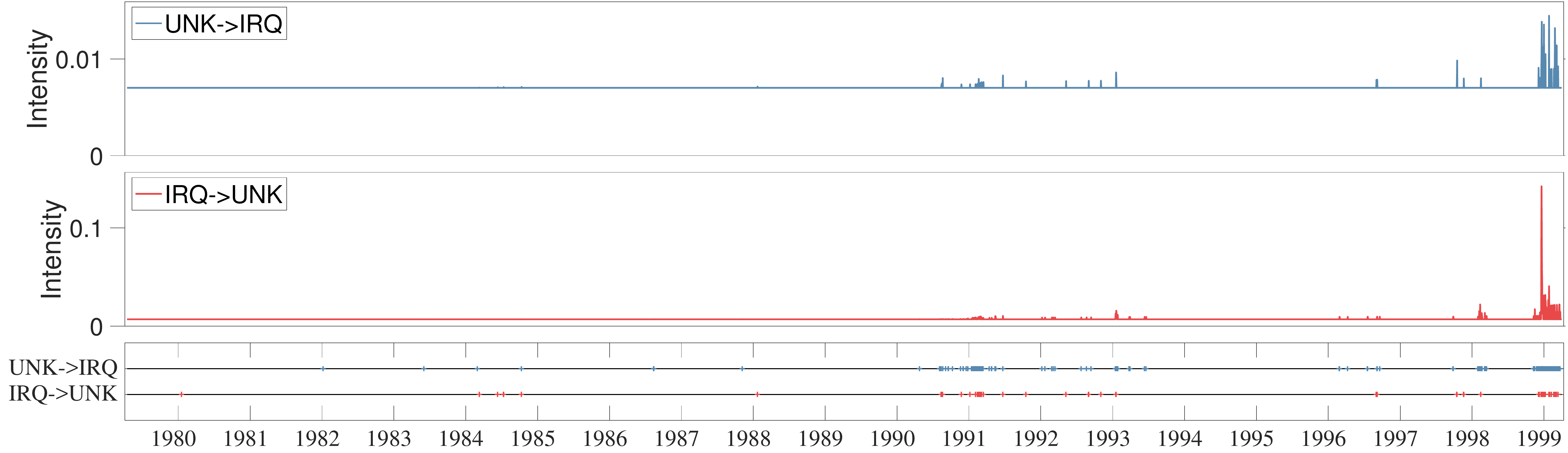}
      \includegraphics[width=14cm,height=15cm, keepaspectratio]{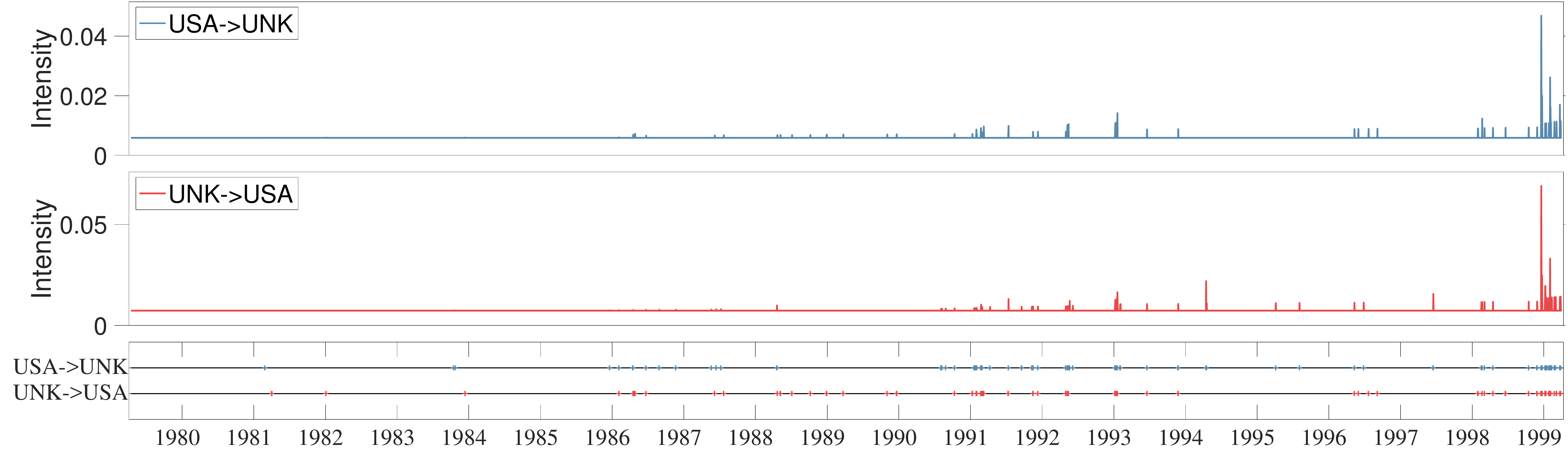}
       \vspace{-0.51em}
  \caption{The plots show the intensity of interaction events among nations inferred by the Hawkes-EPM in the Gulf dataset.}
  \label{figure_intensity23}
\end{figure}

\Cref{figure_intensity21,figure_intensity22,figure_intensity23} present the additional plots of the intensities of the interaction events between the nations: Iran (IRN)-USA, Israel (ISR)-Leban (LEB), Israel (ISR)-Palestin(PAL),Iraq (IRQ)-Israel (ISR), Iraq (IRQ)-Kuwait (KUW), Iraq (IRQ)-Saudi Arabi (SAU), USA-Kuwait (KUW), Iraq (IRQ)-Turkey (TUR),  United Kingdom (UNK)-Iraq (IRQ).

\end{document}